\documentclass[prx,a4paper,portrait,aps,twocolumn,superscriptaddress,nofootinbib,floatfix,10pt]{revtex4-2}
\usepackage[utf8]{inputenc}

\usepackage{amsfonts}
\usepackage{graphicx}
\usepackage[caption=false]{subfig}
\usepackage{amsmath}
\usepackage{amssymb}
\usepackage{hyperref}
\usepackage[bb=px, frak=euler, scr=rsfso]{mathalpha}
\usepackage{xcolor}

\usepackage{mathrsfs}
\usepackage{isomath}
\usepackage{amsthm}
\usepackage{epstopdf}
\usepackage{newtxtext,newtxmath}
\usepackage{dsfont}
\usepackage{physics}
\usepackage{booktabs}
\allowdisplaybreaks[4]
\usepackage{matlab-prettifier}
\usepackage{enumitem} 
\usepackage{esvect}

\usepackage[T1]{fontenc}
\usepackage{orcidlink}

\usepackage{bbold}
\usepackage{mathtools}
\usepackage{soul}
\usepackage{placeins}

\hypersetup{colorlinks,linkcolor=blue!60!black,citecolor=blue!60!black,urlcolor=blue!60!black}

\newcommand{\be}{\begin{equation}}
\newcommand{\bea}{\begin{equation}\begin{aligned}}

\newcommand{\ee}{\end{equation}}
\newcommand{\eea}{\end{aligned}\end{equation}}

\newcommand{\de}{\mathrm{d}}

\newcommand{\aver}[1]{\langle #1 \rangle}

\newcommand{\id}{\mathbb 1}
\newcommand{\idop}{\mathbb 1}
\renewcommand{\tr}{\mathrm{Tr}}

\newcommand{\ham}{H}

\newcommand{\X}{X}

\newcommand{\ignore}[1]{}

\renewcommand{\emph}[1]{{\it #1}}

\def\ph{{\varphi}}
\def\dens{\delta \rho}
\def\mrho{{\bar \rho}}

\def\ndd{\xi}

\def\Npxl{N_p}
\def\L{L}
\def\Dz{\Delta}
\def\Ntro{N_t}

\def\Rvec{\mathbf{X}}

\def\phlatt{\mathfrak{\phi}}
\def\rholatt{\delta \mathfrak{\rho}}
\def\nddisc{\mathfrak{\xi}}

\newcommand{\heal}{\xi_{\rm h}}

\newcommand{\normoderho}{\eta_{\mathfrak{\rho}}}

\newcommand{\normodephi}{\eta_{\mathfrak{\phi}}}

\usepackage{cleveref}
\crefname{equation}{Eq.}{Eqs.}
\crefname{section}{Sec.}{Secs.}
\Crefname{section}{Sec.}{Secs.}
\crefname{subsection}{Sec.}{Secs.}
\Crefname{subsection}{Sec.}{Secs.}
\crefname{subsubsection}{Sec.}{Secs.}
\Crefname{subsubsection}{Sec.}{Secs.}
\crefname{appendix}{App.}{Apps.}
\Crefname{appendix}{App.}{Apps.}
\crefname{figure}{Fig.}{Figs.}
\Crefname{figure}{Fig.}{Figs.}
\crefname{proof}{Proof}{Proofs}
\creflabelformat{proof}{#2proof#3}
\crefname{remark}{Remark}{Remarks}
\crefname{prop}{Proposition}{Propositions}

\usepackage[normalem]{ulem}

\DeclareMathAlphabet{\mathcal}{OMS}{cmsy}{m}{n}

\makeatletter
\def\@bibdoi#1{}
\makeatother

\begin{document}

\title{Thermal Entanglement and Out-of-Equilibrium Thermodynamics in 1D Bose Gases}

\author{Julia Math\'e\,\orcidlink{0009-0002-7403-7044}}
\email{julia.mathe@tuwien.ac.at}
\affiliation{Technische Universit\"{a}t Wien, Atominstitut, Stadionallee 2, 1020 Vienna, Austria}
\affiliation{Vienna Center for Quantum Science and Technology, TU Wien, 1020 Vienna, Austria}
\affiliation{Institute for Quantum Optics and Quantum Information (IQOQI), Austrian Academy of Sciences, Boltzmanngasse 3, 1090 Vienna, Austria}

\author{Nicky Kai Hong Li\,\orcidlink{0000-0002-4087-4744}}
\affiliation{Technische Universit\"{a}t Wien, Atominstitut, Stadionallee 2, 1020 Vienna, Austria}
\affiliation{Vienna Center for Quantum Science and Technology, TU Wien, 1020 Vienna, Austria}
\affiliation{Institute for Quantum Optics and Quantum Information (IQOQI), Austrian Academy of Sciences, Boltzmanngasse 3, 1090 Vienna, Austria}

\author{Pharnam Bakhshinezhad\,\orcidlink{0000-0002-0088-0672}}
\affiliation{Technische Universit\"{a}t Wien, Atominstitut, Stadionallee 2, 1020 Vienna, Austria}
\affiliation{Vienna Center for Quantum Science and Technology, TU Wien, 1020 Vienna, Austria}

\author{Giuseppe Vitagliano\,\orcidlink{0000-0002-5563-3222}}
\email{giuseppe.vitagliano@ific.uv.es}
\affiliation{Technische Universit\"{a}t Wien, Atominstitut, Stadionallee 2, 1020 Vienna, Austria}
\affiliation{Vienna Center for Quantum Science and Technology, TU Wien, 1020 Vienna, Austria}
\affiliation{Instituto de F\'isica Corpuscular, Universitat de Val\'encia -- Consejo Superior de Investigaciones Cient\'ificas, Parc Cient\'ific, E-46980 Paterna, Valencia, Spain}

\begin{abstract}
We investigate entanglement in and out of equilibrium in a one-dimensional Bose gas in its low-energy Bogoliubov regime. In this Gaussian setting, the state is fully characterized by its covariance matrix, which allows us to detect and {\it quantify} entanglement directly from the density and phase fluctuations of the gas. For thermal states, we determine the optimal entanglement witness arising from the covariance matrix criterion and show that it has a remarkably simple form: it depends only on the density fluctuations of the lowest normal mode and on the phase fluctuations of the highest one. For Dirichlet boundary conditions, this gives the exact temperature below which the gas is entangled, which, for many pixels, is given by $k_BT_{\rm sep}\simeq0.83\,\hbar c_0/\Delta$, where $c_0$ is the speed of sound and $\Delta$ the size of the pixels used to resolve the gas. We then study out-of-equilibrium dynamics induced by unitary compression and show that entanglement can be generated even from initially separable thermal states. This requires a non-adiabatic compression, since an adiabatic compression acts separately, and in the same way, on every pixel and therefore conserves entanglement. Moreover, the amount of entanglement created is limited by the compression ratio. We finally discuss how a subsequent contact with a thermal bath, modeled by global or local master equations, affects entanglement. Our methods and results provide a unified and physically intuitive picture of how entanglement emerges and evolves in 1D quantum Bose gases, and identify a simple optimal witness that is relevant more broadly to the analysis of entanglement in quadratic bosonic models and of its role in thermodynamic transformations.
\end{abstract}

\maketitle

\section{Introduction}

Entanglement is one of the most distinctive signatures of genuinely quantum many-body behavior, and its characterization has become a central goal in the study of both equilibrium and out-of-equilibrium systems. Beyond its foundational role, entanglement provides a natural language for quantifying the departure of a many-body state from a classical or mean-field description.
In this sense, entanglement is not only a resource for quantum technologies, but also a diagnostic tool for understanding the structure of complex quantum matter~\cite{amico08,Laflorencie16,girvin_yang_2019}.

A major difficulty, however, is that entanglement is notoriously hard to access in realistic many-body settings, particularly for mixed states and in situations where only a restricted set of observables is experimentally available~\cite{HorodeckiEntanglementReview2009,GuehneToth2009,FriisVitaglianoMalikHuberReview19}. In such cases, one seeks entanglement criteria that are both physically meaningful and operationally accessible. Entanglement witnesses provide a natural framework in this direction: they allow one to certify entanglement from expectation values of suitable observables, without requiring full state tomography~\cite{HorodeckiEntanglementReview2009,GuehneToth2009}. Specifically, these are observables $W$ whose expectation value satisfies $\Tr(W \varrho) <0$ for at least one entangled state $\varrho \in \rm ENT$, while $\Tr (W \sigma) \geq 0$ for all separable states $\sigma \in \rm SEP$.

An especially appealing perspective is to connect entanglement detection to quantities that are central in statistical mechanics and thermodynamics. Over the years, several works have shown that thermodynamically relevant observables, including energy~\cite{Dowling2004,Toth2005,Guhne2005,Guhne2006,Anders_2008,Igloi2022,LiXiMunozAriasReuerHuberFriis2026}, heat capacity~\cite{wiesniak_heat_capacity,Singh_2013}, magnetic susceptibilities~\cite{Brukner2004,Wiesniak2005}, and more general response or thermodynamically relevant functions~\cite{marty14,marty16,CramerPlenioWunderlich11,entanglementbyproxy,Hauke2016,Mazza_2025}, can reveal the presence of entanglement in many-body systems. This line of thought is conceptually attractive because it relates quantum correlations to measurable macroscopic properties and suggests that entanglement may serve as an organizing principle for the interpretation of equilibrium phases and driven dynamics.

It is therefore natural to consider these questions in settings that are both physically relevant and theoretically tractable, such as Gaussian many-body field theories, where entanglement has been studied in the ground state~\cite{audenaert2002,BoteroReznik2004,Cramer_Eisert_et_al_2006,Marcovitch2009,Casini_2009,calabrese_neg_PRL,calabrese2013entanglement,di2020entanglement}, in thermal equilibrium~\cite{Anders_2008,anders2008entanglementseparabilityquantumharmonic,Ferraro_et_al_PRL_2008,Cavalcanti_et_al_2008,Calabrese_2014}, and in certain out-of-equilibrium scenarios, particularly quenches~\cite{Plenio_2004,Eisler_2014,Coser_2014,Hoogeveen_2015,Wen_2015,Angel_Ramelli_2020}. These systems provide an especially convenient setting for formulating entanglement criteria directly in terms of covariance matrices~\cite{werner2001bound,hyllus2006optimal,adesso2007entanglement}, making them readily accessible across a broad range of scenarios. Yet, even in this favorable setting, it remains unclear how complicated the optimal witness structure is, whether entanglement can be certified from only a few observables, and how such a description extends systematically beyond equilibrium.
In this context, low-energy one-dimensional Bose gases are particularly relevant as they admit a Gaussian low-energy description in the Bogoliubov regime, where the relevant degrees of freedom are collective density and phase fluctuations~\cite{PitaevskiiStringari03,cazalillaetalrev11,MoraCastin}, and the state of the system is fully characterized by their covariance matrices.

These questions are timely also from the experimental side. Low-dimensional cold-atom platforms can now prepare and manipulate quantum fields with remarkable control~\cite{RitschRev13,MorschOberthalerRev06,cronin2009optics,BlochRev08}, potentially even allowing the implementation of thermodynamic cycles beyond the classical regime~\cite{Gluza_2021}, while phase and density fluctuations can be accessed directly~\cite{quantumreadout, Murtadho_2025_phasefluct}. Recent experiments have already probed information-theoretic properties such as mutual information and area-law behavior in quantum field simulators, demonstrating that correlation-based diagnostics are within reach~\cite{Tajik_2023}. This raises the broader question of whether one can develop comparably accessible entanglement-based diagnostics that remain meaningful both in equilibrium and during dynamical protocols.

In this work, we address some of these questions by studying the Bogoliubov regime of a one-dimensional Bose gas using a covariance-based framework. We show that the corresponding many-body entanglement admits a remarkably compact description.
Our main contributions are as follows:
\begin{enumerate}
    \item We implement a Gaussian framework for entanglement detection, based on optimal covariance-matrix witnesses, in the discretized model of the gas, i.e., a chain of coupled harmonic oscillators.
    \item For thermal states and adiabatic evolutions, we show that the optimal witness involves only two normal-mode variances, namely the density fluctuations of the lowest mode and the phase fluctuations of the highest one, as indicated, without proof of optimality, in other contexts~\cite{anders2008entanglementseparabilityquantumharmonic,Mazza_2025}, and, for Dirichlet boundary conditions, gives the exact separability temperature, which, for many pixels, is $k_BT_{\rm sep}\simeq0.83\,\hbar c_0/\Delta$, where $c_0$ is the speed of sound and $\Delta$ the pixel size.
    \item We show that non-adiabatic compression can create entanglement even from separable thermal states, within a bound set by the compression ratio, whereas adiabatic compression, which acts in the same way on every pixel, conserves it.
    \item We study the entanglement of the steady states reached with global and with local thermal baths after the compression.
\end{enumerate}

Although we focus on the experimentally motivated study of weakly interacting 1D Bose gases, the methods and results presented here are more general: At low energies, the Bogoliubov theory maps the system onto a collection of coupled harmonic oscillators, and our methods apply to any system described by such a quadratic model. Besides Bose gases, such systems include, e.g., harmonic lattices, trapped-ion phonons, optomechanical arrays, superconducting circuits, and other Gaussian many-body systems. Importantly, the required covariance matrices, and thus the witnesses discussed here, are experimentally accessible in some of these setups~\cite{quantumreadout,Tajik_2023,Murtadho_2025_phasefluct}.
Nevertheless, the Bose gas remains a paradigmatic system of special relevance for investigating thermodynamics in the quantum many-body regime, which is the main focus of our work.

Our work is structured as follows. In \Cref{sec:methods}, we introduce our framework. First, in \Cref{sec:Discretization}, we discretize the Bogoliubov Hamiltonian, and we describe the initial thermal state in \Cref{sec:thermal_state}. Then, we model unitary compression in the discretized model (\Cref{sec:compression_dilation}). In \Cref{sec:ent_wit_meas}, we discuss the covariance matrix criterion as an entanglement quantifier. Our results are presented in \Cref{sec:results} and summarized in \Cref{fig:regimes}. In \Cref{sec:LL_optimal_witness_equilibrium}, we present the optimal witness and the separability temperature of thermal states, and compare them with the numerical results, while the proofs are given in \Cref{app:thermal_witness_proof}. In \Cref{sec:entanglement_under_compression}, we study unitary compression, first in the adiabatic regime (\Cref{sec:LL_optimal_witness_adiabatic}), where the thermal results carry over, and then beyond it (\Cref{sec:nonadiabatic_compression}), where entanglement can be generated within a simple bound set by the compression ratio. In \Cref{sec:thermalization_after_compression}, we discuss the subsequent equilibration with a thermal bath, described by global and local master equations: the global master equation drives the gas to the Gibbs state of the compressed Hamiltonian, whose entanglement follows from our results for thermal states, whereas the local master equation leads to a different steady state, for which we evaluate the optimal witness numerically. Finally, we conclude and present an outlook in \Cref{sec:conclusions}, while technical details are collected in the appendices.

\section{Background and Methods}\label{sec:methods}

\subsection{Discretized description for a 1D Bose gas}\label{sec:Discretization}

\begin{figure}[!tp]
    \centering
    \includegraphics[width=\linewidth]{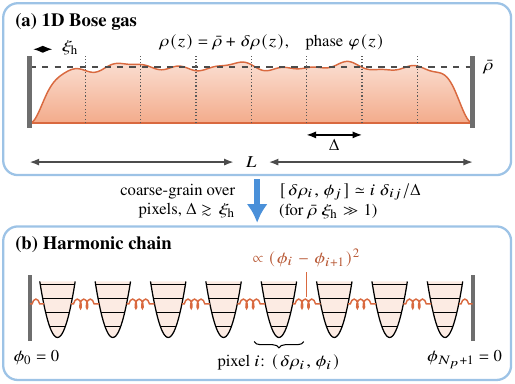}
    \caption{\textbf{Discretizing a 1D BEC.} (a) We start from the Bogoliubov Hamiltonian in Eq.\,\eqref{eq:H_Bogoliubov} that describes a degenerate Bose gas of mean density $\mrho$ in a box of length $L$, in the limit of weak interactions and small density fluctuations. The healing length $\heal=\hbar/\sqrt{mg\mrho}$ is the shortest length scale of this description, e.g., the distance over which the density recovers its bulk value next to a wall or a localized perturbation. (b) We then coarse-grain the phase and density-fluctuation operators over pixels of width $\Delta$, following the procedure from Ref.~\cite{MoraCastin}. Each pixel $i$ becomes a bosonic mode with a canonical pair $(\rholatt_i,\phlatt_i)$, which obeys a rescaled commutation relation, and neighboring pixels are coupled through the phase gradient (springs): the gas becomes a chain of coupled harmonic oscillators, here with the phase pinned at the edges, $\phlatt_0=\phlatt_{\Npxl+1}=0$ (Dirichlet boundary conditions). This also realizes a high-momentum cutoff that can be naturally associated with the healing length of the system, i.e., $\Delta\mathrel{\gtrsim}\heal$.}
    \label{fig:model}
\end{figure}

{\noindent}Our starting point is the Bogoliubov Hamiltonian that approximates a 1D Bose gas in a box of length $L$ and is given by
\be\label{eq:H_Bogoliubov}
\begin{aligned}
 \!\! \ham_\text{B} &= \hspace{-3pt} \int_0^L \hspace{-2pt} \de z \left[ \frac{\hbar^2}{2m} \, \bar \rho \, (\partial_z \ph)^2 + \frac{\hbar^2}{8m \bar \rho} \, (\partial_z  \dens)^2 + \frac g 2  \, \dens^2 \right] ,
\end{aligned}
\ee
where $\bar \rho$ is the mean-field density (fixed by the chemical potential in a grand-canonical ensemble), $\rho = \bar \rho \id + \dens$ is the operator associated with the density, which is composed of a constant mean-field term plus the density-fluctuation operator, and $\ph$ is the phase fluctuation operator. In the following, we set $\hbar = 1$ to simplify the notation, but we restore its value in SI units in the numerical calculations and whenever we give physical values.

Density and phase fluctuations are approximately canonically conjugate, i.e., they obey $[\dens (z),\ph(x)] \simeq i \delta(z-x)  \id$. This holds when there are many atoms per healing length, $\mrho\heal\gg1$, i.e., in the weakly interacting regime (see \Cref{app:bogoliubovtheory}). Density and phase fluctuations thus form a canonical pair at each point in space, analogous to the position and momentum of a particle~\cite{Serafini2017,eisert2010}.
These two canonically conjugate operators arise from the polar decomposition of the atomic field operator $ \Psi^\dagger(z)=\sqrt{ \rho(z)}~e^{-i \ph(z)}$.
See Fig.\,\ref{fig:model} for an overview of the model and \Cref{app:bogoliubovtheory} for further details.

A lattice version of the phononic Hamiltonian in Eq.\,\eqref{eq:H_Bogoliubov} can be obtained by discretizing the interval $[0,\L]$ into $\Npxl$
pixels, each of size $\Dz= L/\Npxl$~\cite{MoraCastin}. This provides a high-momentum cutoff, which is useful for numerical implementations but, as we will see later, is also important as a regularizer of the theory when studying entanglement. For fixed $\Npxl$, the index can be written as $i=1,\ldots, \Npxl+1$, and hence the coordinates of the discretization lattice become $z_i = \frac{L} {\Npxl}(i-1)$. These coordinates then define discretization pixels which are the closed intervals $I_i=[z_i,z_{i+1}]$ for $i=1,\dots,\Npxl$.

We then introduce the discretized versions of the density and phase operators by coarse-graining the field operators via
	\begin{align}
		\phlatt_i &=  {\frac{ 1 }{\Dz} }\int_{I_i} \de z\ \ph( z), \\
	    \rholatt_{i} &= {\frac{ 1 }{\Dz} }\int_{I_i} \de z\  \dens( z),
		\label{eq:x_discrete2}
	\end{align}
with $|I_i|= \Dz= {L}/{\Npxl}$. Writing them in a vector as
	\begin{align}
		\Rvec =\left(\rholatt_1,\ldots,\rholatt_{\Npxl},\phlatt_1,\ldots, \phlatt_{\Npxl}\right)^T,
	\end{align}
we can write their commutation relations compactly as
	\begin{align}\label{eq:resc_CCR_X}
		[ \X_j,  \X_k]= i \frac{\Omega_{j,k}}{\Dz}   \idop ,
	\end{align}
where $\Omega = \left(\begin{smallmatrix}
	    \mathbb{0} & \mathbb{1} \\ -\mathbb{1} & \mathbb{0}
	\end{smallmatrix}\right)$ is the canonical symplectic form.
    That is, their commutation relations are a {\it rescaled} version of the usual \textit{canonical commutation relations} (CCR), with a factor $1/\Delta$.
Note that in the continuum limit $\Npxl \rightarrow\infty$ we have $\Delta \rightarrow 0$ and the discretized operators tend to the continuous density and phase fields with $[\rholatt_{j}, \phlatt_k] = i \delta_{jk} \idop /\Dz \rightarrow [\delta \rho(x), \ph(y)] = i \delta(x-y) \idop$, i.e., the commutation relations correctly tend to the Dirac delta limit.

In summary, to discretize the model, we follow Ref.\,\cite{MoraCastin} (see also Ref.\,\cite{Gluza_2021}) and obtain, in the lowest-order approximation, a discrete quadratic Hamiltonian\footnote{We mostly consider open boundary conditions of Dirichlet type for the phase field, i.e., the phase is pinned at the edges, $\phlatt_0=\phlatt_{\Npxl+1}=0$, since this model would otherwise feature a zero mode. We also tested other boundary conditions, e.g., open boundary conditions of Neumann type (vanishing current through the edges, as for hard walls), and in that case we ``regularize'' the model (i.e., eliminate the zero mode) in the numerical simulations. This can be done in a simple way, while maintaining Gaussianity, by adding a term of the form
\be
 \ham_{\rm reg} = \frac{\Dz}{2}\,\mu \sum_{i=1}^{\Npxl} \phlatt_i^2 .
\ee
We also note that the different boundary conditions do not change our qualitative conclusions, although the simple analytic results of \Cref{sec:results} rely on Dirichlet boundary conditions. A discussion of the relevance and impact of zero modes on entanglement can be found in Ref.\,\cite{aimet2026compactnesscurbsentanglementgrowth}.
Note also that, with respect to Eq.\,\eqref{eq:H_Bogoliubov}, i.e., the full Bogoliubov model, we are neglecting the second term, which contains the spatial derivative of the density fluctuations, and in the discretized
model would take the form~\cite{MoraCastin}
\be
    \ham_{\rm bog} = -\frac{\Dz}{8m \mrho } \sum_{i=2}^{\Npxl-1} \frac{\rholatt_i \, (\rholatt_{i+1} +\rholatt_{i-1} -2\rholatt_i)}{\Dz^2} .
\ee
This would still fit into the Gaussian framework that we describe below. However, it is a first non-universal correction term which makes the numerical calculations
more computationally costly and less accurate.
}:
\begin{align}\label{eq:discretHamilt}
      \ham &\approx \Dz \left[\frac{\mrho}{2m} \sum_{i=0}^{\Npxl} \left( \frac{ \phlatt_i-\phlatt_{i+1}}{\Dz}\right)^2 + \frac{g} 2 \sum_{i=1}^{\Npxl} \rholatt_i^2 \right] ,
\end{align}
where, for Dirichlet boundary conditions, $\phlatt_0=\phlatt_{\Npxl+1}=0$. Formally, this model is analogous to a chain of coupled harmonic oscillators (see Fig.\,\ref{fig:model}).

\subsection{Description of initial thermal state}\label{sec:thermal_state}

{\noindent}Throughout this work, we take the initial state to be the Gibbs state of the
discretized Hamiltonian,
\begin{equation}
    \tau = \frac{e^{-\ham/T}}{\mathcal{Z}},
    \qquad \text{with} \quad
    \mathcal{Z}=\tr\left(e^{-\ham /T}\right),
\end{equation}
where we set $k_B=1$.
Since $\ham$ is quadratic in bosonic quadratures, $\tau$ is a Gaussian state and is therefore fully characterized by its first and second moments.
In particular, in our case the first moments vanish, i.e., $\aver{\X_i}=0$, and the only non-trivial object is the covariance matrix:
\begin{equation}\label{eq:CM_def}
    (\Gamma_\tau)_{ij} =
    \,\aver{\X_i \X_j + \X_j \X_i}_\tau
    \;-\;
    2 \aver{\X_i}_\tau \aver{\X_j}_\tau.
\end{equation}
Since we work in the Heisenberg picture, expectation values are taken with respect to the initial Gibbs state $\tau$, and we will often omit its label to simplify the notation.
Compactly, the discretized Hamiltonian can be written in standard quadratic form as
\begin{equation}\label{eq:H_quadratic_form}
    H
    \;\approx\;
    \frac{\Dz}{2}\,\Rvec^{\,T}
    \begin{pmatrix}
        H_{\mathfrak{\rho}} & 0\\[2pt]
        0 & H_{\mathfrak{\phi}}
    \end{pmatrix}
    \Rvec ,
\end{equation}
with
\begin{equation}\label{eq:XP_matrices_LL}
    H_{\mathfrak{\rho}} = g\,\id,
    \qquad
    H_{\mathfrak{\phi}} = \frac{\mrho}{m\,\Dz^2}\,\mathbb L ,
\end{equation}
where $\mathbb L$ is the discrete Laplacian associated with the gradient term
in Eq.~\eqref{eq:discretHamilt}, which for Dirichlet boundary conditions includes the boundary terms $(\phlatt_0-\phlatt_1)^2=$$\phlatt_1^2$ and $(\phlatt_{\Npxl}-\phlatt_{\Npxl+1})^2=$$\phlatt_{\Npxl}^2$. Since the pinned phases $\phlatt_0$ and $\phlatt_{\Npxl+1}$ are not dynamical variables, $\mathbb L$ is an $\Npxl\times\Npxl$ matrix, all of whose diagonal entries are equal to $2$ (see \Cref{app:discretized_model}).

In the present case, the thermal covariance matrix is block diagonal,
\begin{equation}\label{eq:Gamma_blockdiag}
    \Gamma_{\tau}
    =
    \begin{pmatrix}
        \Gamma_{\mathfrak{\rho}}(T) & 0\\
        0 & \Gamma_{\mathfrak{\phi}}(T)
    \end{pmatrix},
\end{equation}
with the same block structure as the Hamiltonian.
Both the covariance matrix and the Hamiltonian can be decomposed into normal-mode quadratures in the following way (see also \Cref{app:discretized_model_dynamics}). Starting with the Hamiltonian, one finds the orthogonal matrix $O$ that diagonalizes $H_{\mathfrak{\phi}}$,
and uses it to define the \textit{normal-mode quadratures}\footnote{For Dirichlet boundary conditions, the columns of $O$ are standing (discrete sine) waves, see \Cref{app:discretized_model}: the lowest mode, $k=1$, has the smoothest profile across the whole gas, whereas the profile of the highest mode, $k=\Npxl$, alternates in sign from one pixel to the next.},
\begin{equation} \label{eq:orthonormalmodetransf}
    \boldsymbol{\eta}_{\mathfrak{\rho}} = O^T \boldsymbol{\delta \rho},
    \qquad
    \boldsymbol{\eta}_{\mathfrak{\phi}} = O^T \boldsymbol{\phlatt}.
\end{equation}
In the normal-mode basis, the Hamiltonian becomes
\begin{equation}
   H
   = \Dz
   \sum_{k=1}^{\Npxl}
   H_k,
   \quad \text{with} \quad
   H_k
   =
   \frac12
   \Big(
   g\,(\normoderho)_k^2
   +
   \nu_k\,(\normodephi)_k^2
   \Big),
\end{equation}
where $\nu_k$ are the eigenvalues of $H_{\mathfrak{\phi}}$.

The thermal second moments are then given by
\begin{align}\label{eq:mode_covariances}
    \aver{(\normoderho)_k^2}_\tau
    &
   = \frac{\omega_k}{2 g \Delta }\, \coth \left(\frac{\omega_k}{2T}\right),
    \\
    \aver{(\normodephi)_k^2}_\tau
    &=
    \frac{g}{2\omega_k\Delta}\,
    \coth \left(\frac{\omega_k}{2T}\right) ,\label{eq:mode_covariances2}
\end{align}
where $\omega_k =\sqrt{g \nu_k}$ are the normal-mode frequencies, which grow with $k$. In the following, we denote by $\vartheta_k:=\coth(\omega_k/2T)=2n_k+1$ the occupation factor of mode $k$, where $n_k$ is its mean thermal occupation, i.e., the mean number of phonons in the mode. See \Cref{app:discretized_model} for further details.

\subsection{Compression as a time-dependent dilation}
\label{sec:compression_dilation}

\begin{figure*}[!t]
    \centering
    \includegraphics[width=1.0\textwidth]{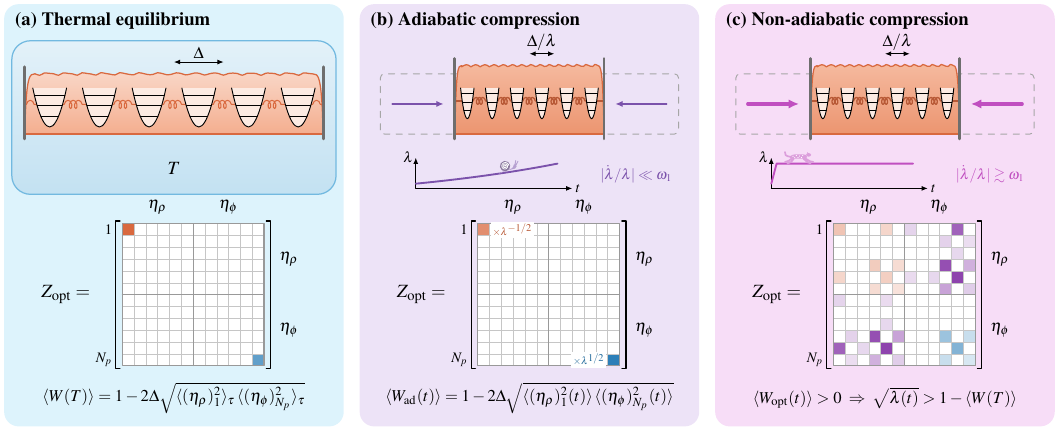}
    \caption{\textbf{Entanglement in the three regimes studied in this work.} Top: the gas as a chain of coupled oscillators on a uniform condensate [cf.\ Fig.~\ref{fig:model}(b)]. (a)~Contact with a bath at temperature $T$. (b),(c)~Unitary compression to $L/\lambda$  with the exponential protocol $\lambda(t)$ of \Cref{sec:compression_dilation}, slow (b) or fast (c) compared
    with the lowest mode, Eq.~\eqref{eq:adiabaticity_condition}: the pixels shrink to $\Delta/\lambda$, the density grows, and the couplings stiffen. Bottom: optimal CMC witness matrix $|Z_{\rm opt}|$, Eq.~\eqref{eq:linear_witness_rewrite}, in the normal-mode basis;    orange, blue and purple cells are density--density, phase--phase and density--phase  entries, each block normalized separately (density and phase carry different units), on a common scale for all panels.    (a)~For thermal states only two entries remain, the density of mode~1 and the phase of mode~$N_p$, giving Eq.~\eqref{eq:optwit_thermal_analytic} (\Cref{sec:LL_optimal_witness_equilibrium}). (b)~An adiabatic compression acts as the same local squeezing on every pixel, Eq.~\eqref{eq:adiabatic_covariance_local_squeezing}: $Z_{\rm opt}$ keeps its two-entry form with entries rescaled by $\lambda^{\mp1/2}$ (\Cref{app:local_squeezing_invariance}), and the witness is conserved, Eq.~\eqref{eq:adiabatic_witness}. (c)~A non-adiabatic compression squeezes and rotates each mode differently; $Z_{\rm opt}$ spreads over several modes and acquires density--phase entries. Entanglement can be created only if $\sqrt{\lambda(t)}>1-\langle W(T)\rangle$, Eq.~\eqref{eq:compression_bound}. Matrices: SDP of \Cref{app:OptSDPwit} for $N_p=6$, starting from the thermal state at $T_{\rm sep}$, with $\lambda_f=2$; in (c), $|\dot\lambda/\lambda|=30\,\omega_1\simeq7\,\omega_{N_p}$,    shown at $(t-t_{\rm comp})\,\omega_1\simeq3.07$, where $\langle W_{\rm opt}\rangle\simeq0.08$.}
    \label{fig:regimes}
\end{figure*}

{\noindent}Let us now model a time-dependent length change
(see Fig.~\ref{fig:regimes}(b),(c) and also Ref.\,\cite{Gluza_2021}).
Let $L(t)$ be the time-dependent length of the box.
A convenient parametrization of the compression/expansion is via the ratio
\begin{equation}
\lambda(t):=\frac{L(0)}{L(t)},
\end{equation}
so that compression corresponds to increasing $\lambda$.
A convenient way to account for the changing system size is to switch to a co-compressing coordinate $z\mapsto x :=\lambda(t)\,z,$ such that
\begin{equation}
\begin{aligned}
z&=\frac{x}{\lambda(t)}, \quad
\partial_z =\lambda(t)\,\partial_x, \quad
\de z=\frac{\de x}{\lambda(t)} ,
\end{aligned}
\end{equation}
so that the spatial domain is fixed to $x\in[0,L]$ with $L=L(0)$ at all times (see \Cref{app:compression}).
Since the canonical commutator involves a Dirac delta, the coordinate dilation must be accompanied by a
field rescaling~\cite{Gluza_2021}. Physically, compression squeezes the same atoms into a shorter interval, so that the density fluctuations per unit length grow by a factor $\lambda$, while the phase, being dimensionless, is not rescaled.
Indeed, using $\delta(z-z')=\lambda(t)\delta(x-x')$, one finds that defining
\begin{equation}
\label{eq:nu_def}
\begin{aligned}
\ndd(x,t)\; &:=\;\frac{1}{\lambda(t)}\,\dens \left(\frac{x}{\lambda(t)},t\right), \\
\ph(x,t)\; &:= \ph \left(\frac{x}{\lambda(t)},t\right),
\end{aligned}
\end{equation}
preserves the canonical structure in the fixed domain:
\begin{equation}
\big[\ndd(x,t),\ph(x',t)\big]= i\,\delta(x-x').
\end{equation}
In this co-compressing frame, the time-dependent Hamiltonian (before discretization) becomes~\cite{Gluza_2021}
\begin{equation}
\label{eq:H_LL_cocompressing}
 H(t)=\int_{0}^{L}\!\!\de x\,
\Bigg[
\frac{\mrho(0)}{2m}\,\lambda^2(t)\,(\partial_x \ph)^2
+\frac{g}{2}\,\lambda(t)\,\ndd^{\,2}
\Bigg],
\end{equation}
where we used $\mrho(t)=N/L(t)=\mrho(0)\lambda(t)$.

The co-compressing frame formulation is particularly convenient for the discretized model with coarse-grained fields
$\phlatt_i$ and $\rholatt_i$.
In fact, in the laboratory frame, implementing a compression $L(t)$ at fixed $\Npxl$ requires a time-dependent lattice spacing
\be
\Dz(t)=L(t)/\Npxl = \Dz / \lambda(t) \ ,
\ee
where we called $\Dz \equiv \Dz(0)$ the initial lattice spacing.
Defining the discrete co-compressing density fluctuations
\begin{equation}
\label{eq:nu_i_def}
\nddisc_i(t):=\frac{1}{\lambda(t)}\,\rholatt_i(t),
\end{equation}
while leaving the phases $\phlatt_i(t)$ unchanged, one ensures that the canonical algebra is time-independent in the fixed-domain discretization, i.e.,
\be
[\nddisc_j(t) , \phlatt_k(t)] = i \delta_{jk} / \Dz .
\ee

Furthermore, we discretize the time interval, such that the overall length change $L \mapsto L + \delta L$ is implemented in $\Ntro$ small steps.
Calling $t_{\rm comp}$ the total compression time, we divide it into $\Ntro$ steps of duration
$
\delta t=t_{\rm comp}/\Ntro
$.
At step $n$ (with $n=0,\ldots,\Ntro-1$), the total length is $L_n$, which we update iteratively as
\begin{equation}
L_{n+1}=(1+\epsilon)L_n ,
\label{eq:L_step}
\end{equation}
with
$\epsilon$ fixed by $(1+\epsilon)^{\Ntro}=1+\delta L/L$, so that the final length is exactly $L+\delta L$; for $\Ntro\gg1$, $|\epsilon|\ll 1$.

Thus, at step $n$ (with $n=0,\ldots,\Ntro-1$), the total length and the corresponding compression factor are
\[
L_n=L_0(1+\epsilon)^n
\quad \text{and} \quad
\lambda_n=\frac{L_0}{L_n}=(1+\epsilon)^{-n},
\]
with $\epsilon$ being the relative length change per step (negative for compression).
In the continuous-time limit, this corresponds to the exponential protocol $\lambda(t)=\lambda_f^{t/t_{\rm comp}}$, with final compression factor $\lambda_f:=(1+\epsilon)^{-\Ntro}$ and constant rate $|\dot\lambda/\lambda|=|\ln\lambda_f|/t_{\rm comp}$.
Further details on the compression dynamics and on its numerical implementation are given in \Cref{app:compression}.

Note that the passage from the physical lattice density fluctuations $\rholatt_i(t)$ to the co-compressing variables
$\nddisc_i(t)$ in Eq.\,\eqref{eq:nu_i_def} corresponds to a purely local, time-dependent canonical rescaling, i.e., $\textbf{Y}(t) = R(t) \textbf{X} (t)$, where $\textbf{Y}(t) := (\nddisc_1(t), \dots , \nddisc_{\Npxl}(t), \phlatt_1(t) , \dots , \phlatt_{\Npxl}(t))^T$ is the vector of co-compressing pixel variables and
$R(t):=\big(\lambda^{-1}(t)\,\mathbb{1}_{\Npxl}\big)\oplus \mathbb{1}_{\Npxl}$.
Since this transformation acts on each pixel separately, it cannot generate or destroy entanglement: the laboratory and the co-compressing descriptions contain exactly the same entanglement between pixels. In the following, a tilde denotes covariance matrices expressed in terms of the co-compressing variables, i.e., $\tilde\Gamma(t)$ is defined as in Eq.~\eqref{eq:CM_def}, with $\textbf{X}$ replaced by $\textbf{Y}(t)$.

\subsection{Covariance matrix criterion as an entanglement quantifier}\label{sec:ent_wit_meas}

{\noindent}It is well known that, especially for Gaussian states, entanglement in bosonic systems can be certified and quantified directly from the covariance matrix.
One typical example is the so-called {\it entanglement negativity}~\cite{vidalwerner2002,PlenioPRL2005}, which can be calculated from the symplectic eigenvalues of the covariance matrix after partial transposition of one subsystem (cf.\,\Cref{app:ent_measures}).
Another paradigmatic and powerful approach is the \emph{covariance matrix criterion} (CMC)~\cite{werner2001bound,hyllus2006optimal}, which works as follows in a general $\Npxl$-mode system.
Let $\Gamma_\varrho$ denote the covariance matrix of a state $\varrho$ with respect to a canonical operator vector
$\textbf{X} =(x_1,\ldots, x_{\Npxl},p_1,\ldots,p_{\Npxl})^T$ satisfying the CCR, $[X_j , X_k] = i \Omega_{jk} \id$, with symplectic form $\Omega$.
If $\varrho$ is separable across a bipartition $A|B$, i.e., $\varrho = \sum_k q_k (\varrho_A \otimes \varrho_B)_k$ with $q_k\geq0$ and $\sum_k q_k=1$, then there exist \emph{local} covariance matrices $\gamma_A$ and $\gamma_B$ such that
\be\label{eq:CMC_AB_rewrite}
    \Gamma_\varrho - \gamma_A \oplus \gamma_B \succeq 0,
\ee
where $\gamma_A$ and $\gamma_B$ satisfy the Heisenberg uncertainty relation on their respective subsystems,
\be\label{eq:CMC_AB_HUP_rewrite}
    \gamma_A + i\Omega_A \succeq 0,
    \qquad
    \gamma_B + i\Omega_B \succeq 0 .
\ee
Equivalently, if no such $\gamma_A,\gamma_B$ exist (i.e., if Eq.\,\eqref{eq:CMC_AB_rewrite} is violated), then $\varrho$ is necessarily entangled across $A|B$.

Clearly, the discussion about the entanglement of a many-mode state will depend on the choice of bipartition, and certain partitions
can exhibit more entanglement than others. In this context, it is worth recalling the connection with \emph{area-law} behavior of entanglement~\cite{eisert2010}. For ground states of local, gapped many-body Hamiltonians, entanglement across a bipartition is often expected to scale primarily with the size of the \emph{boundary} between the two regions, rather than with their volume. In harmonic and more general bosonic lattice models, rigorous area-law results are known for non-critical finite-range systems, and closely related bounds also exist for thermal correlations, for instance in terms of the mutual information~\cite{cramer2006,amico08,wolf2008area,eisert2010}. Hence, when comparing different bipartitions of a spatially local system, one generally expects the amount of detectable entanglement to depend strongly on the geometry of the cut: bipartitions with a larger interface can support larger boundary entanglement~\cite{Ferraro_et_al_PRL_2008,Cavalcanti_et_al_2008}, whereas in one-dimensional gapped settings the entanglement across a connected cut typically remains bounded~\cite{audenaert2002}. This provides useful intuition for why some partitions are more favorable than others when assessing the total amount of entanglement.

The same idea naturally extends to partitions into more than two parts. In our case, we consider \emph{fully separable} states of $\Npxl$ modes, i.e., states that can be decomposed as
\be
\varrho=\sum_k q_k\,(\varrho_1\otimes\cdots\otimes \varrho_{\Npxl})_k .
\ee
The CMC can be extended to this scenario and requires the existence of single-mode covariance matrices $\gamma_j$ such that
\be\label{eq:CMC_full_rewrite}
    \Gamma_\varrho - \gamma_1 \oplus \cdots \oplus \gamma_{\Npxl} \succeq 0,
\ee
with $\gamma_j+i\Omega\succeq 0$ for each mode $j$.
For general (non-Gaussian) states, this provides a necessary separability condition, whereas for Gaussian states, the CMC becomes both \emph{necessary} and \emph{sufficient} \cite{hyllus2006optimal}. In our setting, this means that as long as the dynamics is generated by Hamiltonians (and master equations) that are quadratic in the mode operators, separability is fully characterized at the level of the covariance matrix\footnote{Note, however, that this would no longer apply in the presence of higher-order interactions, i.e., beyond the Bogoliubov approximation, which could lead to entanglement invisible to the covariance matrix.}.

The matrix inequality in \cref{eq:CMC_full_rewrite} is equivalent to a family of scalar inequalities obtained by testing against any positive semidefinite matrix $Z\succeq 0$,
\be\label{eq:CMC_linear_family_rewrite}
    \tr\, (Z\,\Gamma_\varrho) \;\ge\; \min_{\{\gamma_j\}}\tr\, \Big(Z\,(\gamma_1\oplus\cdots\oplus\gamma_{\Npxl})\Big).
\ee
The optimization over admissible local covariances (subject to $\gamma_j+i\Omega\succeq 0$) can be formulated as a semidefinite program (SDP)~\cite{hyllus2006optimal}, see also \Cref{app:OptSDPwit}. Importantly, the dual SDP yields a positive semidefinite matrix $Z_{\mathrm{opt}}$ which defines an entanglement witness via
\be\label{eq:linear_witness_rewrite}
    \aver{W}_\varrho \;:=\; 1 - \tr\, (Z_{\mathrm{opt}}\Gamma_\varrho) ,
\ee
which is non-positive for all fully separable states, while certifying entanglement whenever it is positive\footnote{Note that we adopt a sign convention opposite to the standard definition of entanglement witness, essentially taking the negative of the witness as an entanglement quantifier.}.
Diagonalizing $Z_{\mathrm{opt}}=\sum_\ell \Lambda_\ell\,\mathbf V_\ell \mathbf V_\ell^\dagger$, one obtains a criterion in the form of an uncertainty relation for variances:
\be
    \tr( Z_{\mathrm{opt}} \,\Gamma_\varrho)=2\sum_\ell \text{Var} (\tilde X_\ell)_\varrho,
    \label{eq:witness_variances}
\ee
where $\tilde X_\ell=\sqrt{\Lambda_\ell}\,\mathbf V_\ell^\dagger\cdot\mathbf X$ are linear combinations of quadratures and $\text{Var} (\tilde X_\ell)_\varrho = \aver{\tilde X^2_\ell}_\varrho - \aver{\tilde X_\ell}^2_\varrho$ are their variances calculated on the quantum state $\varrho$.
Thus, CMC witnesses can be interpreted as uncertainty relations for suitable collective quadratures.

At this point, it is important to remember that in our case we work with variables that are canonical but rescaled. Hence, instead of simply applying Eqs.\,\eqref{eq:CMC_AB_rewrite} and \eqref{eq:CMC_full_rewrite} with the constraints $\gamma_j + i \Omega \succeq 0$, we must consider the appropriately rescaled versions, i.e.,
\be\label{eq:CMC_full_rewrite_rescaled}
    \Gamma_\varrho - \gamma_1 \oplus \cdots \oplus \gamma_{\Npxl} \succeq 0,
    \quad \gamma_j + i \frac{1}{\Delta} \Omega \succeq 0.
\ee
For more details, see \Cref{app:OptSDPwit}.

\section{Results}\label{sec:results}

{\noindent}Our results cover the three regimes sketched in Fig.~\ref{fig:regimes}(a)--(c). In thermal equilibrium (\Cref{sec:LL_optimal_witness_equilibrium}), we find the optimal witness, which involves only two normal modes, and a simple formula for the exact separability temperature in terms of physical parameters. Under adiabatic compression (\Cref{sec:LL_optimal_witness_adiabatic}), the entanglement of the initial thermal state is exactly conserved. Under non-adiabatic compression (\Cref{sec:nonadiabatic_compression}), entanglement can instead be created, by an amount that is limited by the compression ratio. Finally, in \Cref{sec:thermalization_after_compression}, we discuss what happens when the compressed gas is brought into contact with a thermal bath.

\subsection{Entanglement and optimal witness of thermal states}
\label{sec:LL_optimal_witness_equilibrium}

\begin{figure}[!tbp]
    \centering
    \includegraphics[width=\linewidth]{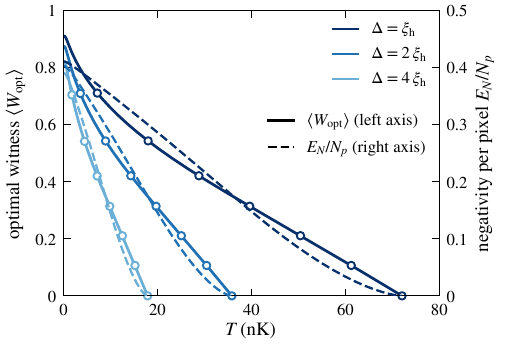}
    \caption{Entanglement of thermal states as a function of temperature, for a gas of length $L=50\,\mu$m with $6128$ $^{87}$Rb atoms and three pixel sizes at or above the healing length, $\Delta=\heal\simeq0.25\,\mu$m, $2\,\heal$ and $4\,\heal\simeq1.0\,\mu$m ($\Npxl=197$, $98$ and $49$ pixels). Optimal CMC witness (solid lines, left axis): analytic result, Eq.~\eqref{eq:optwit_thermal_analytic}, and numerical optimum of the SDP (circles). Logarithmic negativity of the alternating bipartition $A=\{1,3,5,\ldots\}$, $B=\{2,4,6,\ldots\}$, per pixel (dashed lines, right axis). Only the entangled region is shown: both quantities vanish at the same temperature $T_{\rm sep}$, where the curves end ($T_{\rm sep}=72.1$, $35.8$ and $17.9$\,nK), in agreement with Eq.~\eqref{eq:Tsep_estimate}.}
    \label{fig:ent_over_temp}
\end{figure}

We first apply the covariance matrix criterion (CMC) to the Gibbs state $\tau$ of the discretized
Hamiltonian, whose normal-mode variances are given
in Eqs.~\eqref{eq:mode_covariances} and~\eqref{eq:mode_covariances2}. We use
the SDP formulation of the CMC described above to find the optimal witness against full
separability, and we plot it as a function of temperature, for typical values of the parameters of the model, in \Cref{fig:ent_over_temp}.

We also compare the CMC witness with the logarithmic negativity and with CMC witnesses optimized separately for each bipartition. Among the bipartitions that we tested, the alternating cut
$A=\{1,3,5,\ldots\}$, $B=\{2,4,6,\ldots\}$ gives the strongest signal, in agreement with known results and with the area-law behavior mentioned in \Cref{sec:ent_wit_meas}~\cite{audenaert2002,Ferraro_et_al_PRL_2008,Cavalcanti_et_al_2008}, and it detects
entanglement whenever the full-separability witness does. In fact, for Dirichlet boundary conditions, we show below that its negativity and the full-separability witness vanish exactly at the same temperature.

The numerical optimum has a remarkably simple form, which we can also derive analytically (the proofs are given in \Cref{app:thermal_witness_proof}). In thermal equilibrium,
the normal mode frequencies $\omega_k$ grow with $k$, and Eqs.~\eqref{eq:mode_covariances} and~\eqref{eq:mode_covariances2} show that the density fluctuations $\langle(\normoderho)_k^2\rangle_\tau$ grow with $k$, while the phase fluctuations $\langle(\normodephi)_k^2\rangle_\tau$ decrease with $k$. Thus, the lowest mode has the smallest density fluctuations, and the highest mode has the smallest phase fluctuations. The optimal witness involves only these two collective variables,
\begin{equation}
    (\normoderho)_1=\sum_{j=1}^{\Npxl}u_j\,\rholatt_j ,
    \qquad
    (\normodephi)_{\Npxl}=\sum_{j=1}^{\Npxl}w_j\,\phlatt_j ,
    \label{eq:extremal_collective_quadratures}
\end{equation}
where $u_j=O_{j1}$ and $w_j=O_{j\Npxl}$ are the spatial profiles of the lowest and of the highest normal mode, respectively. The two variables commute, because they belong to different normal modes. Nevertheless, their fluctuations cannot both be small in a separable state: every fully separable state $\sigma$, Gaussian or not, obeys (cf.~\cref{app:witness_proof})
\begin{equation}
    \operatorname{Var}[(\normoderho)_1]_\sigma\,\operatorname{Var}[(\normodephi)_{\Npxl}]_\sigma
       \geq\frac{C_{1\Npxl}^{\,2}}{4\Delta^2},
    \label{eq:general_product_sep_condition}
\end{equation}
where $C_{1\Npxl}:=\sum_{j}|u_jw_j|\leq1$ is the overlap of the two mode profiles on the pixels. This bound follows from the uncertainty relation that each pixel obeys individually, $[\rholatt_j,\phlatt_j]=i/\Delta$, and a violation of Eq.~\eqref{eq:general_product_sep_condition} therefore certifies entanglement.

For Dirichlet boundary conditions, the highest mode is the lowest one with alternating signs, $w_j=(-1)^{j+1}u_j$, so that $C_{1\Npxl}=1$ (see \Cref{app:discretized_model}). In this case, the criterion~\eqref{eq:general_product_sep_condition} is also sufficient for the thermal state: whenever it is satisfied, the thermal state can be written explicitly as a mixture of product states (\Cref{app:thermal_witness_proof}).
The thermal state is thus entangled if and only if the corresponding witness, i.e., the witness of Eq.~\eqref{eq:linear_witness_rewrite} built only from the two variables of Eq.~\eqref{eq:extremal_collective_quadratures}, with optimal weights,
\begin{equation}
    \begin{aligned}
    \langle W(T)\rangle
     &=1-2\Delta\sqrt{\langle(\normoderho)_1^2\rangle_\tau\langle(\normodephi)_{\Npxl}^2\rangle_\tau}\\
     &=1-\sqrt{
        \frac{\omega_1}{\omega_{\Npxl}}
        \coth\!\left(\frac{\omega_1}{2T}\right)
        \coth\!\left(\frac{\omega_{\Npxl}}{2T}\right)},
    \end{aligned}
    \label{eq:optwit_thermal_analytic}
\end{equation}
is positive. This witness is also the optimum of the SDP at every temperature (\Cref{app:thermal_witness_proof}), which we also observe numerically.
Criteria based on the lowest and highest normal modes have been discussed before~\cite{anders2008entanglementseparabilityquantumharmonic,Mazza_2025}. Here, we prove that, for Dirichlet boundary conditions, the corresponding witness is optimal at every temperature and gives the exact separability threshold. For other boundary conditions, the overlap $C_{1\Npxl}$ is smaller than one, e.g., $C_{1\Npxl}\to2\sqrt2/\pi\approx0.90$ for the Neumann chain at large $\Npxl$. The criterion of Eq.~\eqref{eq:general_product_sep_condition} then still certifies entanglement, but the exact threshold has to be obtained from the full SDP (\Cref{app:thermal_witness_proof}).

For $T=0$, this witness reduces to $\langle W\rangle=1-\sqrt{\omega_1/\omega_{\Npxl}}$, which for Dirichlet boundaries depends only on $\Npxl$. At temperatures $T\gg\omega_1\simeq\pi c_0/L$, where $c_0=\sqrt{g\mrho/m}$ is the speed of sound, we can approximate $\coth(\omega_1/2T)\simeq2T/\omega_1$, and the witness depends only on the ratio $y:=\omega_{\Npxl}/2T$ between the energy of the highest mode and the temperature, as $\langle W\rangle\simeq1-\sqrt{\coth(y)/y}$ (see \Cref{app:analytic_scaling}). The witness thus vanishes when $\coth y_*=y_*$, i.e., at $y_*\simeq1.20$. Using $\omega_{\Npxl}\simeq2c_0/\Delta$ and restoring physical units, we obtain a simple formula for the temperature below which the gas is entangled,
\begin{equation}
    k_BT_{\rm sep}\simeq0.42\,\hbar\omega_{\Npxl}\simeq0.83\,\frac{\hbar c_0}{\Delta}=0.83\,\frac{\heal}{\Delta}\,g\mrho .
    \label{eq:Tsep_estimate}
\end{equation}
Hence, the separability temperature
depends only on the speed of sound and on the pixel size, and not on the size of the gas. Entanglement between pixels of the size of the healing length
persists up to temperatures of the order of the chemical potential $g\mrho$, i.e., up to the boundary of the regime described by our model.
Conversely, at a given temperature $T$, entanglement between pixels can only be certified for pixel widths smaller than $\Delta_{\max}\simeq0.83\,\hbar c_0/(k_BT)$.
The formal limit $\Delta\to0$ would send $\langle W(T)\rangle\to1$ at any temperature (\Cref{app:continuum_limit}). Physically, however, the pixel size has to be consistent with the regime of validity of our approximation: since we work in the phononic regime, we consider pixels at least of the order of the healing length, $\Delta \mathrel{\gtrsim} \heal$ (see \Cref{app:continuum_limit}).

Entanglement is thus certified from two normal-mode
fluctuations, in particular when they are jointly smaller than allowed by the uncertainty relation of single pixels. Experimentally, both quantities are variances of fixed linear combinations of pixel densities and pixel phases, e.g., $\operatorname{Var}[(\normoderho)_1]=\mathbf u^T\Gamma_{\mathfrak{\rho}}\mathbf u/2$, where $\mathbf u=(u_1,\ldots,u_{\Npxl})^T$ and $\Gamma_{\mathfrak{\rho}}$ is the density block of the covariance matrix. Phase profiles of elongated and 1D Bose gases are routinely extracted from matter-wave interference after time of flight~\cite{Dettmer_2001,Manz_2010,Murtadho_2025_phasefluct}, and the density fluctuations can be reconstructed from the subsequent dynamics~\cite{quantumreadout}, so that the required covariance matrices are within reach of current experiments~\cite{Tajik_2023}. The two variances are static two-point correlation functions, namely density--density correlations at the longest wavelength and phase--phase correlations at the shortest wavelength resolved by the pixels, similar to the spin--spin correlations probed, e.g., by neutron scattering in magnetic materials~\cite{Hauke2016,Laurell2024}. The main requirement is that the pixel size $\Delta$ is resolved, since the highest mode varies on the scale of single pixels.

Finally, the witness also quantifies entanglement. Every witness of the form of Eq.~\eqref{eq:linear_witness_rewrite}, normalized such that $\tr(Z\Gamma_\sigma)\geq1$ for all fully separable states $\sigma$, gives a lower bound on an entanglement measure called {\it best separable approximation} (BSA)~\cite{LewSanp98,KarnasLew01},
\begin{equation}
    E_{\rm BSA}^{\mathcal S}(\varrho)
      \geq\max\{0,1-\tr(Z\Gamma_\varrho)\},
    \label{eq:EBSAbound}
\end{equation}
which quantifies the smallest weight $p$ of an arbitrary remainder component in a decomposition of the form $\varrho = (1-p) \sigma + p \varpi$ where $\sigma$ is a separable state (see \Cref{app:bsa}). In particular, the thermal state satisfies $E_{\rm BSA}^{\mathcal S}(\tau)\geq\max\{0,\langle W(T)\rangle\}$. The BSA thus measures how far $\varrho$ is from the set of separable states, similarly to the geometric measure of entanglement~\cite{WeiGoldbart2003}. More generally, the measured value of a witness can be used to bound other convex entanglement measures~\cite{G_hne_2007,eisert2007quantitative}. The same two modes also control the logarithmic negativity of the alternating cut: for Dirichlet boundaries, it is nonzero if and only if $\langle W(T)\rangle>0$, in which case $E_N\geq-\log_2[1-\langle W(T)\rangle]$ (see \Cref{app:ent_measures}). This explains why the two quantifiers in \Cref{fig:ent_over_temp} vanish at the same temperature.

\subsection{Entanglement under unitary compression}
\label{sec:entanglement_under_compression}

We now study how the witness changes when the gas is compressed unitarily, i.e., without contact with a bath. Since the Hamiltonian remains quadratic, the state remains Gaussian, and it suffices to follow its covariance matrix in time. We show that adiabatic protocols conserve the initial entanglement, so that the analysis of \Cref{sec:LL_optimal_witness_equilibrium} carries over, whereas faster protocols can create entanglement.

We describe the compression in the co-compressing pixel variables $\mathbf Y(t)$, which contain the same entanglement between pixels as the laboratory ones. The key simplification is that the compression only multiplies the density and phase terms of the Hamiltonian, Eq.~\eqref{eq:H_LL_cocompressing}, by $\lambda(t)$ and $\lambda^2(t)$, respectively. For Dirichlet boundary conditions, the normal modes therefore keep their shape at all times, and only their parameters change. In terms of the normal-mode variables $(\normoderho)_k=\sum_jO_{jk}\nddisc_j$ and $(\normodephi)_k=\sum_jO_{jk}\phlatt_j$, which coincide at $t=0$ with those of the initial thermal state, the Hamiltonian splits into a sum of contributions from independent oscillators as in
\begin{equation}
    \begin{gathered}
    H(t)=\Delta\sum_{k=1}^{\Npxl}\frac12\left[g\lambda(t)\,(\normoderho)_k^2+\lambda^2(t)\,\nu_k\,(\normodephi)_k^2\right],
    \end{gathered}
    \label{eq:compression_mode_Hamiltonian}
\end{equation}
with the oscillators vibrating at frequencies $\omega_k(t)=\lambda^{3/2}(t)\,\omega_k$, where $\nu_k$ and $\omega_k=\sqrt{g\nu_k}$ are the values at $t=0$. All frequencies thus grow by the same factor $\lambda^{3/2}(t)$: the speed of sound grows as $\lambda^{1/2}$ because the density grows as $\lambda$, while all wavelengths shrink as $1/\lambda$. Since these normal modes evolve independently, they never become correlated.

Intuitively, the compression is adiabatic if $\lambda(t)$ varies slowly compared with the oscillation of every mode. Since the lowest mode oscillates most slowly, this requires
\begin{equation}
    \left|\frac{\dot\lambda(t)}{\lambda(t)}\right|\ll\omega_1(t)\simeq\frac{\pi c(t)}{L(t)} ,
    \label{eq:adiabaticity_condition}
\end{equation}
where $c(t)=\sqrt{g\mrho(t)/m}$ is the instantaneous speed of sound. Physically, the compression must be slow compared with the time a phonon needs to cross the gas. For the exponential protocol, $|\dot\lambda/\lambda|=|\ln\lambda_f|/t_{\rm comp}$, so that Eq.~\eqref{eq:adiabaticity_condition} requires $t_{\rm comp}\gg|\ln\lambda_f|/\omega_1$. Note that this is a condition on the physical protocol, unrelated to the time step used in the numerical propagation.

\subsubsection{Adiabatic compression}
\label{sec:LL_optimal_witness_adiabatic}

If Eq.~\eqref{eq:adiabaticity_condition} holds, each mode follows its slowly changing Hamiltonian, and its mean number of phonons $n_k$ does not change. The variances of each mode are then given by Eqs.~\eqref{eq:mode_covariances} and~\eqref{eq:mode_covariances2} with $g\to g\lambda$ and $\nu_k\to\lambda^2\nu_k$, but with the \emph{initial} occupation factor $\vartheta_k=\coth(\omega_k/2T)$. Hence, the density and phase fluctuations of every mode are rescaled by the same factors,
\begin{equation}
    \begin{aligned}
    \langle(\normoderho)_k^2(t)\rangle_{\tau}&=\lambda^{1/2}(t)\,\langle(\normoderho)_k^2(0)\rangle_{\tau} ,\\
    \langle(\normodephi)_k^2(t)\rangle_{\tau}&=\lambda^{-1/2}(t)\,\langle(\normodephi)_k^2(0)\rangle_{\tau} ,
    \end{aligned}
    \label{eq:adiabatic_density_variance}
\end{equation}
and no correlations between density and phase build up. Since all frequencies grow by the same factor $\lambda^{3/2}(t)$, while the occupations do not change, the compressed gas is precisely in the Gibbs state of the compressed Hamiltonian $H(t)$ at the higher temperature $T_{\rm eff}(t)=\lambda^{3/2}(t)\,T$: like a classical gas, the phonon gas heats up when it is compressed adiabatically.

For entanglement, the crucial point is that all modes are rescaled by the \emph{same} factors $\lambda^{\pm1/2}$. In terms of the co-compressing pixel variables $\mathbf Y(t)$,
the adiabatic state is therefore obtained from the initial thermal state by squeezing every pixel in the same way,
\begin{equation}
    \begin{gathered}
    \tilde\Gamma_{\rm ad}(t)=D_\lambda(t)\Gamma_\tau D_\lambda^T(t),
    \end{gathered}
    \label{eq:adiabatic_covariance_local_squeezing}
\end{equation}
with $D_\lambda(t) =\lambda^{1/4}(t)\mathbb{1}_{\Npxl} \oplus\lambda^{-1/4}(t)\mathbb{1}_{\Npxl}$.
Note that here $D_\lambda(t)$ relates the covariance matrix of $\mathbf Y(t)$ to that of $\mathbf Y(0)=\Rvec(0)$,
and it is a symplectic transformation, namely the same single-mode squeezing on every pixel~\footnote{It should not be confused with the change of variables $\mathbf Y(t)=R(t)\Rvec(t)$ of \Cref{sec:compression_dilation}, which rescales only the densities and is thus not a squeezing transformation. Equation~\eqref{eq:adiabatic_covariance_local_squeezing} is already written in the co-compressing variables $\mathbf Y(t)$. In terms of the laboratory variables $\Rvec(t)$, the covariance matrix of the adiabatic state is $R^{-1}(t)D_\lambda(t)\,\Gamma_\tau\,D_\lambda^T(t)R^{-T}(t)$, with $R^{-1}(t)D_\lambda(t)=\lambda^{5/4}(t)\mathbb{1}_{\Npxl}\oplus\lambda^{-1/4}(t)\mathbb{1}_{\Npxl}$, which also acts on each pixel separately.}. Since $D_\lambda(t)$ acts on each pixel separately, it can neither create nor destroy entanglement: \emph{adiabatic compression conserves the initial entanglement}. In other words, the compression heats the gas, but it raises the separability temperature by the same factor $\lambda^{3/2}$, since all mode frequencies grow by this factor. In particular, the two-mode witness keeps its value, because the product of the two variances in Eq.~\eqref{eq:general_product_sep_condition} does not change,
\begin{equation}
    \begin{aligned}
    \langle W_{\rm ad}(t)\rangle_{\tau}
       &=1-2\Delta\sqrt{\langle(\normoderho)_1^2(t)\rangle_{\tau}\,\langle(\normodephi)_{\Npxl}^2(t)\rangle_{\tau}}\\
       &=\langle W(t=0)\rangle_\tau ,
    \end{aligned}
    \label{eq:adiabatic_witness}
\end{equation}
for any compression ratio. Here, the subscript $\tau$ recalls that the expectation values of the time-dependent (Heisenberg-picture) observables are taken with respect to the initial thermal state, and $\langle W(t=0)\rangle_\tau$ is the witness of that state, denoted by $\langle W(T)\rangle$ in Eq.~\eqref{eq:optwit_thermal_analytic}.
More generally, a witness of the initial state, transformed by the same local squeezing, is still a valid witness, with the same value, and the transported two-mode witness is also the optimum of the SDP for the adiabatically compressed state (\Cref{app:local_squeezing_invariance}), as we also verified numerically.

\subsubsection{Beyond the adiabatic regime}
\label{sec:nonadiabatic_compression}

For faster protocols, the compression also squeezes each mode, by an amount that depends on its frequency (see Fig.~\ref{fig:regimes}(c)): this creates correlations between the density and the phase of each mode and, since it no longer acts in the same way on all modes, it can create entanglement between the pixels.
To compute the witness for arbitrary protocols, we propagate the covariance matrix in the co-compressing frame
(cf.~\Cref{app:compression}, see also Refs.~\cite{DupaysPRB2024,DupaysPRB2025} for related
finite-time dynamics in driven Tomonaga--Luttinger models), and at every selected time we solve the SDP of \Cref{app:OptSDPwit}. As a benchmark, we also evaluate the initial optimal witness, transported by the local squeezing $D_\lambda$ of Eq.~\eqref{eq:adiabatic_covariance_local_squeezing}. Since $D_\lambda$ acts on each pixel separately, this transported witness remains valid for any protocol.

\begin{figure}[!tbp]
    \centering
    \includegraphics[width=\linewidth]{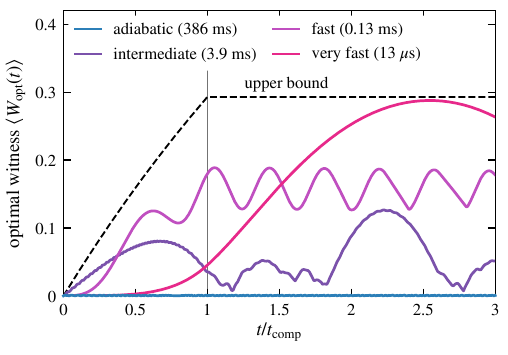}
    \caption{Optimal CMC witness during a compression of the gas to half its length, $\lambda_f=2$, with $\lambda(t)=\lambda_f^{t/t_{\rm comp}}$, for four compression times $t_{\rm comp}$ (in parentheses). After the compression (right of the vertical line), the length is kept fixed and the gas is isolated. The initial state is the thermal state at the separability threshold, $T_0=T_{\rm sep}=72.1$\,nK, i.e., $\langle W(T_0)\rangle=0$. Adiabatic compression ($|\dot\lambda/\lambda|=0.01\,\omega_1$): the witness stays at zero (within $\sim\!10^{-3}$), obeying Eq.~\eqref{eq:adiabatic_witness}, and no entanglement is created. Intermediate compression ($|\dot\lambda/\lambda|=\omega_1$): creates weak entanglement, $\langle W_{\rm opt}\rangle\lesssim0.13$. Fast compression ($|\dot\lambda/\lambda|=30\,\omega_1$): the witness reaches $0.19$ shortly after the compression, below the upper bound $1-1/\sqrt{\lambda(t)}$ of Eq.~\eqref{eq:compression_bound} (dashed line), which is approached only by compressions that are fast also compared with the oscillations of the highest mode (\Cref{app:compression_bound}). For the very fast compression ($|\dot\lambda/\lambda|=300\,\omega_1$), the witness reaches $0.288$ at $t\simeq2.5\,t_{\rm comp}$, close to the bound, $0.293$. Here, $\Delta=\heal\simeq0.25\,\mu$m ($\Npxl=197$, $\omega_1=180\,{\rm s}^{-1}$) and all other parameters are those of Fig.~\ref{fig:ent_over_temp}.}
    \label{fig:compression_regimes}
\end{figure}

\Cref{fig:compression_regimes} illustrates the different regimes for a gas that is compressed to half its length, starting from the thermal state at the separability threshold, which is thus not entangled. The adiabatic compression leaves the witness unchanged, as predicted by Eq.~\eqref{eq:adiabatic_witness}. The intermediate compression, with $|\dot\lambda/\lambda|=\omega_1$, lies at the border of the adiabatic regime defined by Eq.~\eqref{eq:adiabaticity_condition}: the witness then changes less as compared to the fast compression case. Such small changes can create entanglement only if the initial state is very close to the separability threshold, as in this example. The fast compression, $|\dot\lambda/\lambda|=30\,\omega_1$, instead creates much more entanglement, which keeps growing for a short time after the compression, while the gas is isolated at fixed length. An even faster compression, $|\dot\lambda/\lambda|=300\,\omega_1\simeq2.4\,\omega_{\Npxl}$, which is fast also compared with the oscillations of the highest mode, creates almost the largest entanglement allowed by the compression ratio, see Eq.~\eqref{eq:compression_bound} below. Since an adiabatic compression conserves entanglement, the creation of entanglement  also signals a departure from adiabaticity, and hence the performance of irreversible work. At the same time, away from the adiabatic regime the transported witness can be far from optimal, and solving the full SDP is needed: after the intermediate compression, for instance, the transported witness (not shown) is negative most of the time, whereas the optimal one is almost always positive.

Finally, the entanglement that a compression can create is limited by the compression ratio, as we can observe heuristically as follows.
In the co-compressing variables, the shrinking of the pixels $\Delta(t)=\Delta/\lambda(t)$ is absorbed into the rescaled densities $\nddisc_j$ and only modifies the parameters of the Hamiltonian (\Cref{sec:compression_dilation}). An adiabatic compression follows this change by rescaling the fluctuations of every pixel in the same way, as in Eq.~\eqref{eq:adiabatic_covariance_local_squeezing}, which is why the transported thermal witness remains optimal. In a faster compression, instead, the fluctuations (and hence also the optimal witness) follow the same rescaling as the pixels. Quantitatively, we can consider
a sequence of small sudden steps, in which the pixels shrink as $\lambda\to\lambda+\de\lambda$. In such a sudden step, the covariance matrix of the co-compressing variables does not change, while the adiabatic state, instead, does change: its density variances grow and its phase variances shrink by the same factor $1+\de\lambda/(2\lambda)$. Each step can therefore make the fluctuations of the gas smaller than those of the adiabatic state at most by this factor, and all together the factors of all steps multiply to $\lambda^{1/2}(t)$. The free evolution between the steps does not change this comparison, since the adiabatic state is a thermal, and thus stationary, state of the instantaneous Hamiltonian. As a consequence, the variance of any linear combination of pixel densities and phases is, at all times, at least $\lambda^{-1/2}(t)$ times its value in the adiabatic state (cf.~\Cref{app:compression_bound}). Since the adiabatic state has the same optimal witness as the initial thermal state, i.e., $\tr(Z\tilde\Gamma_{\rm ad})\geq1-\langle W(T)\rangle$ for every normalized witness matrix $Z$, the optimal witness obeys
\begin{equation}
    \langle W_{\rm opt}(t)\rangle_{\tau}\leq1-\frac{1-\langle W(T)\rangle}{\sqrt{\lambda(t)}} .
    \label{eq:compression_bound}
\end{equation}
A more detailed derivation of this bound is given in \cref{app:compression_bound}.
This bound is approached, for many pixels, shortly after a sudden compression, and it shows that a compression can create entanglement from a separable thermal state only if the temperature is close enough to $T_{\rm sep}$, within a window set by the compression ratio (\Cref{app:compression_bound}). For instance, reducing the length by $10\%$ can create at most $\langle W\rangle\approx0.05$, and only for initial temperatures up to about $7.5\%$ above $T_{\rm sep}$, while a gas at twice the separability temperature can only become entangled if its length is reduced by a factor of about three or more.

\subsection{Local and global equilibration with thermal bath}
\label{sec:thermalization_after_compression}

\begin{figure}[!tbp]
    \centering
        \includegraphics[width=\linewidth]{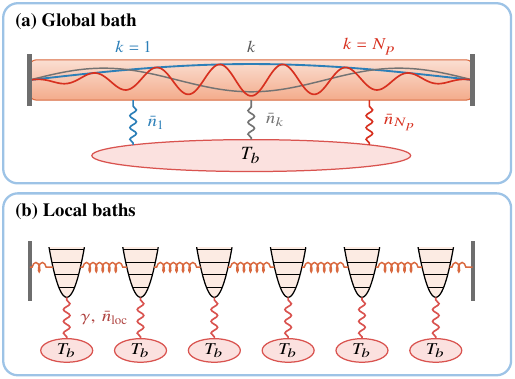}
    \caption{\textbf{Global and local baths.} (a) A global bath couples to the normal modes of the compressed gas, and each mode $k$ relaxes to its own thermal occupation $\bar n_k$ at the bath temperature $T_b$. The steady state is thus the Gibbs state of $H_f$, which is entangled if and only if $T_b<T_{\rm sep}^f=\lambda_f^{3/2}T_{\rm sep}$. (b) Local baths act on each pixel separately, with the same damping rate $\gamma$ and the same thermal occupation $\bar n_{\rm loc}$.
    Their noise is not matched to the normal modes, so that the steady state is in general not a Gibbs state (\Cref{fig:local_steady_state}).}
    \label{fig:baths_graphics}
\end{figure}

We now consider the dynamics of the gas when it is brought into
contact with a thermal bath at temperature $T_b$ and fixed length
$L_f$ (see \Cref{fig:baths_graphics}). In the weak system-bath coupling regime~\cite{Breuer2007},
the open-system dynamics is governed by the
Gorini--Kossakowski--Sudarshan--Lindblad (GKSL) master equation,
which contains a Hamiltonian evolution which
is quadratic and jump operators that are linear in the mode operators (see \Cref{app:thermal_after_comp}).
Thus, the
dynamics is Gaussian-preserving, and the full many-body evolution
can be tracked at the level of the covariance matrix $\Gamma(t)$.

Specifically (cf.~\Cref{app:thermal_after_comp}), the GKSL equation reduces
to a Lyapunov equation for the covariance matrix
\begin{equation}
    \dot\Gamma = G\Gamma + \Gamma G^T + D,
\end{equation}
where the drift matrix $G$ combines coherent evolution under
$H_f$ and dissipative damping, and the diffusion matrix $D$
encodes the fluctuations injected by the bath. Since $G$ and
$D$ are time-independent,
this equation can be integrated exactly. Given the steady-state
covariance matrix $\Gamma_\infty$, defined by $\dot\Gamma=0$,
the covariance matrix at any time $t$ is
\begin{equation}
    \Gamma(t) = \Gamma_\infty + e^{Gt}\bigl[\Gamma(0)
    - \Gamma_\infty\bigr]e^{G^T t},
    \label{eq:thermalization_exact_transient}
\end{equation}
where $\Gamma(0)$ is the state at the end of the compression
stroke. The specific forms of $G$, $D$, and $\Gamma_\infty$
depend on the choice of jump operators, as we now discuss.

\begin{figure}[!tbp]
    \centering
        \includegraphics[width=\linewidth]{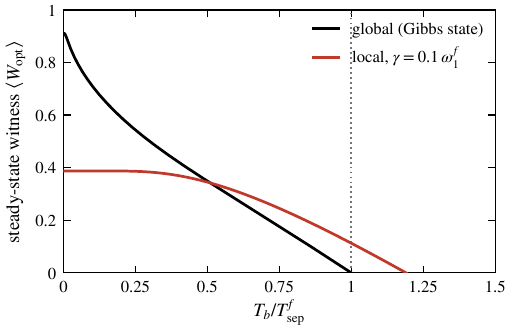}
    \caption{Steady states of global and of local baths, at fixed length after the compression shown in \Cref{fig:compression_regimes} ($\lambda_f=2$, $T_{\rm sep}^f=\lambda_f^{3/2}T_{\rm sep}=204$\,nK). Witness of the steady states as a function of the bath temperature $T_b$; the curves end where entanglement is lost. Black: global bath, whose steady state is the Gibbs state of $H_f$, entangled for $T_b<T_{\rm sep}^f$. Red: local baths (phenomenological local master equation) with weak damping, $\gamma=0.1\,\omega_1^f$; for smaller $\gamma$, the curve does not change. The local steady state is entangled up to $T_b\approx1.19\,T_{\rm sep}^f$, i.e., above the threshold of the Gibbs state. The dotted line marks $T_b=T_{\rm sep}^f$.}
    \label{fig:local_steady_state}
\end{figure}

\textit{Global master equation.} In the global model~\cite{Hofer_2017,
De_Chiara_2018, Gonzalez_2017}, the jump operators are the normal-mode
annihilation operators,
and the bath drives each mode $k$ to
the thermal occupation $\bar{n}_k = [\exp(\omega_k^f/T_b)-1]^{-1}$.
The drift and diffusion matrices are
\begin{equation}
    G = \Omega\mathbf{H}_f - \frac{\gamma}{2}\mathbb{1},
    \qquad
    D_{\rm glob} = \gamma\,\Gamma_\infty^{\rm glob},
\end{equation}
where $\mathbf{H}_f$ is the matrix of quadratic coefficients of $H_f$ and $\Gamma_\infty^{\rm glob}$
is the covariance matrix of the Gibbs state. The structure of
$D_{\rm glob}$ is precisely what guarantees that the unique
steady state is the Gibbs state of $H_f$ at temperature $T_b$~\cite{Toscano_2022} (cf.~\Cref{app:thermal_after_comp}).
Its entanglement as a function
of $T_b$ follows exactly from the discussion in \cref{sec:LL_optimal_witness_equilibrium}.
In particular, for Dirichlet boundary conditions, compression rescales all normal-mode frequencies uniformly as $\omega_k^f = \lambda_f^{3/2}\omega_k$, which raises the separability threshold of the Gibbs state to $T_{\rm sep}^f = \lambda_f^{3/2}T_{\rm sep}$ (black curve in Fig.~\ref{fig:local_steady_state}). Since the steady state is unique, it is independent of $\Gamma(0)$ and retains no memory of the compression.

\textit{Local master equation.} In the local model~\cite{Gonzalez_2017, De_Chiara_2018}, the jump operators
act on individual pixels,
and every pixel is driven toward the same thermal occupation
$\bar{n}_{\rm loc} = [\exp(\omega_{\rm loc}/T_b)-1]^{-1}$,
set by a local reference frequency $\omega_{\rm loc}$. The
drift matrix $G$ is identical to the global case, while the
diffusion matrix,
\begin{equation}
    D_{\rm loc} = \frac{\gamma(2\bar{n}_{\rm loc}+1)}{\Delta}
    \left(\ell^{-1}\mathbb{1}_{N_p}\oplus\ell\,\mathbb{1}_{N_p}\right),
\end{equation}
is diagonal and uniform across all pixels, determined in terms of a fixed length scale $\ell$.
Since $D_{\rm loc}$ does not depend on the
mode index $k$, unlike $D_{\rm glob}$, it cannot encode the
mode-dependent thermal occupations $\bar{n}_k$, and the steady
state of the local master equation is in general not the Gibbs
state of $H_f$. The analytical solution for the local steady
state $\Gamma_\infty^{\rm loc}$ is given in
\Cref{app:thermal_after_comp}.

 Both master equations rely on a weak coupling to the bath, so that $\gamma$ must be small compared with the frequencies of the gas~\cite{Breuer2007} and, for the global master equation, also compared with their spacings (\Cref{app:thermal_after_comp}). The steady state of the global master equation does not depend on $\gamma$, and for the local baths we therefore consider $\gamma \ll \omega_1^f$ (the weak-damping regime), in which the collective modes are only weakly
perturbed by the local baths, and the entanglement threshold
can lie above $T_{\rm sep}^f$. We note, however, that the local master equation in principle also requires the coupling between neighboring pixels $\kappa$ to be weak compared with $\gamma$, whereas in our model, this does not hold since the magnitude of $\kappa$ is comparable to that of the on-site frequency [see Eqs.\,\eqref{eq:app_XP_matrices} and \eqref{eq:app_Laplacian_D}].
Our results should then be understood within a simple phenomenological model.
The optimal witness is evaluated numerically on the
analytical steady state $\Gamma_\infty^{\rm loc}$ and the
results are shown in Fig.\,\ref{fig:local_steady_state}. At intermediate times, the
covariance matrix evolves from the compressed state toward
$\Gamma_\infty$ on a timescale $\propto1/\gamma$, see Eq.~\eqref{eq:thermalization_exact_transient}, so that the entanglement created by the compression is eventually replaced by that of the steady state.

In \Cref{fig:local_steady_state}, we compare the two steady states at the same bath temperature. They behave qualitatively in the same way for cold baths, for which both are entangled, and for hot baths, for which both are separable. They differ in a window of bath temperatures above $T_{\rm sep}^f$, up to $T_b\simeq1.19\,T_{\rm sep}^f$, in which only the local steady state is entangled, whereas for $T_b\lesssim0.5\,T_{\rm sep}^f$ the local steady state is less entangled than the Gibbs state. In this phenomenological comparison, the entanglement of the steady state thus depends on how the bath is modeled.

\section{Conclusions and outlook}\label{sec:conclusions}

{\noindent}In this work, we have investigated how entanglement can be certified and quantified in equilibrium and out-of-equilibrium states of a discretized one-dimensional Bose gas. Our approach combines two ingredients: a discretized Gaussian description of the low-energy field theory in terms of covariance matrices, and the optimization of covariance-matrix-based entanglement witnesses. This provides an operational and physically transparent route to entanglement detection in 1D Bose gases in terms of the density and phase fluctuations of the normal modes. From an entanglement point of view, it is noteworthy that the optimal witness also quantifies entanglement via monotones,
and that its value depends on the level of coarse-graining.

For the initial thermal state, and more generally under fully adiabatic evolutions, we found that the optimal witness, and correspondingly the separability threshold,
depend only on two variances, namely the density fluctuations of the lowest normal mode and the phase fluctuations of the highest one.
This also leads to a simple formula for the separability temperature in terms of physical parameters, $k_BT_{\rm sep}\simeq0.83\,\hbar c_0/\Delta$, set only by the speed of sound and by the size of the pixels.

Another physical result of our analysis is that unitary compression can generate detectable entanglement even when the initial thermal state is separable. In this sense, the compression stroke does not merely reshape the spectrum of excitations, but can actively drive the system across a separability threshold. This requires a non-adiabatic compression, which squeezes different normal modes in different ways, whereas an adiabatic compression acts as the same local squeezing on every pixel and conserves the initial entanglement. The amount of entanglement that a compression can create is limited by the compression ratio as shown in Eq.~\eqref{eq:compression_bound}, and we evaluate the optimal witness in this regime numerically.
When the compressed gas is then coupled to a thermal bath, its entanglement is eventually set by the steady state of the dynamics. In the weak-coupling regime, the global master equation drives the gas to the Gibbs state of the compressed Hamiltonian, which is entangled for $T_b<\lambda_f^{3/2}T_{\rm sep}$ and retains no memory of the compression. A phenomenological local master equation, instead, leads in general to a steady state that is not a Gibbs state and that, for weak damping, can be entangled also above $\lambda_f^{3/2}T_{\rm sep}$ (\Cref{fig:local_steady_state}).
Overall, these findings suggest that entanglement may serve as a sensitive diagnostic of the interplay between coherent, possibly irreversible, driving and dissipative relaxation.

Experimentally, the optimal witness can in principle be evaluated whenever the covariance matrix can be (partially) reconstructed. In the context of 1D Bose gases, this can be done in the phase--density basis~\cite{quantumreadout,Murtadho_2025_phasefluct}, and, for thermal and adiabatically compressed states, the witness only requires the two variances discussed in \Cref{sec:LL_optimal_witness_equilibrium}. An open experimental question is how well the highest normal mode can be measured with a limited imaging resolution.

These observations open up a broader perspective, namely the relation between entanglement and thermodynamic processes. Understanding whether, when, and in what precise sense entanglement plays an operational role in quantum thermodynamics remains an active and only partially understood problem \cite{Binder2018, Campbell_2026, Goold_2016}. This is especially true in genuinely many-body settings, where collective effects and dissipation can become relevant simultaneously \cite{Gluza_2021,Cangemi2024,Mukherjee2024}.

In this context, an important open question is whether the generated entanglement can play an operational role, such as enhancing work extraction or cycle efficiency~\cite{Khandelwal2020, Heineken2020, Diotallevi_2024,OliveiraJunior2024_HeatWitness}, or whether it should instead be regarded primarily as a by-product of many-body dynamics. Recent works have shown that many-body effects, including interactions and entanglement, can influence the work statistics and the performance of quantum heat engines, both in theory~\cite{Boubakour2023,Zawadzki2024,Li2024EntanglementWorkStatistics} and in experiments with ultracold atoms~\cite{Koch2023}. Our witness, which requires only two variances for thermal and adiabatically driven states, offers a practical way to monitor entanglement in such engines and to compare it with independently measured thermodynamic quantities.

\begin{acknowledgments}
{\noindent}We thank Yuri Minoguchi, Taufiq Murtadho, Bi Hong Tiang, Nelly H.Y. Ng, and Marek Gluza, for insightful discussions. This research was funded in whole or in part by the Austrian Science Fund (FWF)  [\href{https://doi.org/10.55776/P35810}{10.55776/P35810}], [\href{https://doi.org/10.55776/P36633}{10.55776/P36633}] and [\href{https://doi.org/10.55776/P36478}{10.55776/P36478}].
For open access purposes, the authors have applied a CC BY public copyright license to any author-accepted manuscript version arising from this submission.
G.V. also acknowledges support from the Grant No. RYC2024-048278-I funded by MCIU/AEI/10.13039/501100011033 and FSE+.
N.K.H.L.~also acknowledges financial support from the Croucher Foundation.
We further acknowledge support from the Austrian Federal Ministry of Education, Science and Research via the Austrian Research Promotion Agency (FFG) through the project FO999921415 (Vanessa-QC) funded by the European Union{\textemdash}NextGenerationEU, and from the European Research Council (Consolidator grant `Cocoquest' 101043705).
The authors used ChatGPT~5.6 and Claude Opus~5.5 for assistance with coding, graphics, symbolic calculations, and the editing of the text.
The authors provided the main ideas and proof directions and checked all results and take full responsibility for the content of the article.
\end{acknowledgments}

\appendix

\section{The Bogoliubov approximation of a 1D BEC}\label{app:bogoliubovtheory}

{\noindent}We consider the Bose gas in one dimension with delta-function interaction, which is the well-known Lieb-Liniger model in a box potential of length $L$. See for example Refs.\,\cite{PitaevskiiStringari03,cazalillaetalrev11}. In second quantization, the Hamiltonian is
\begin{align}\label{eq:LLHam}
 H_\text{LL} = \hspace{-3pt} \int_0^L \hspace{-2pt} \de z \left[ -\frac{\hbar^2}{2m}  \Psi^\dagger \partial_z^2  \Psi + \frac{g} 2  \Psi^\dagger | \Psi|^2  \Psi \right] \ ,
\end{align}
where $ \Psi(z)$ is the field operator that annihilates a particle at position $z$, $m$ is the atomic mass, and $g=2\hbar^2/m a$
is the effective 1D interaction strength that depends on the 1D scattering length $a$\footnote{The 1D scattering length is related to the 3D scattering length $a_{\rm 3D}$ by the relation $a=a^2_\perp/a_{\rm 3D}$, where $a_\perp=\sqrt{\hbar/m \omega_\perp}$ is the radial oscillator length arising from the trapping potential \cite{PitaevskiiStringari03}.}.
As in the main text, here we set our units to $\hbar = 1$, but we keep $\hbar$ explicit in some physical parameters.
As usual, the commutation relation of the field operators is $[ \Psi(z) ,  \Psi^\dagger(z^\prime)]= \delta(z-z^\prime)$. In the hydrodynamic approach, these field operators are written in the polar decomposition,
\begin{equation}
	 \Psi^\dagger(z)=\sqrt{  \rho(z)}~e^{-i \ph(z)} =\sqrt{ \mrho(z)  \id + \dens(z)}~e^{-i \ph(z)},
\end{equation}
thereby introducing a phase operator $\ph$ and also expanding the density operator as $ \rho = \mrho(z)  \id + \dens(z)$, where $\mrho(z)= \aver{ \rho(z)}$ is the mean density profile, which, in the homogeneous case, reads
$\mrho(z) = \tfrac N L$.
In this picture, the relevant operators are the density-fluctuation operator $\dens(z)$ and the phase operator $\ph(z)$, whose commutation relations follow from $e^{i \ph(x)}  \rho(z) e^{-i \ph(x)} -  \rho(z) = \delta(z-x)$, which implies that
\be
[\rho(z),e^{i \ph(x)}]=[\dens(z),e^{i \ph(x)}] = -e^{i \ph(z)} \delta(z-x) .
\ee
This, in turn, implies that in the limit $\mrho\heal\gg1$, the density and phase fluctuation fields are canonically conjugate, i.e., they obey $[ \dens(z),\ph(x)]= i \delta(z-x)  \id$. In this representation, the operator $\ph$ plays the role of the fluid velocity potential, and the fluid velocity operator can be defined as
\be
 v:=\tfrac \hbar m \partial_z \ph ,
\ee
which brings the equations of motion into a form analogous to usual fluid equations.
Note that this description is valid only for large densities, corresponding to
$mg / \hbar^2 \mrho \ll 1$, i.e., the weakly interacting regime of the model \eqref{eq:LLHam}, which is equivalent to having many atoms per {\it healing length}, $\mrho\heal\gg1$, where the latter is defined as
\be \label{eq:healing_length}
\heal = \frac{\hbar}{\sqrt{mg\mrho}} = \frac{\hbar}{mc_0} ,
\ee
and sets the characteristic length scale of the model.

In this representation, and considering only the lowest orders in the fluctuation operators, the Hamiltonian can be written as
\be\label{eq:H_BogoliubovApp1}
  H_\text{B} = \hspace{-3pt} \int_0^L \hspace{-2pt} \de z \left[ \frac{\hbar^2}{2m}\mrho (\partial_z \ph)^2 + \frac{\hbar^2}{8m \mrho(z)} (\partial_z  \dens)^2 + \frac g 2  \dens^2 \right] ,
\ee
which represents the
Bogoliubov approximation to the 1D Bose gas.

The Bogoliubov Hamiltonian is quadratic in the conjugate fluctuation operators, and can thus be diagonalized by a Bogoliubov transformation
\be
 H_\text{B} = E_0 + \sum_k \epsilon_k  b^\dagger_k  b_k ,
\ee
leading to the famous Bogoliubov energy spectrum
\be\label{eq:Bogospectrum}
\epsilon_k = \sqrt{\tfrac{\hbar^2 k^2}{2m} \left( \tfrac{\hbar^2 k^2}{2m} + 2g\mrho \right)} ,
\ee
which is such that for low momenta $k \rightarrow 0$, the dispersion relation becomes linear: $\epsilon_k \simeq \hbar c_0 k$ where $c_0=\sqrt{g\mrho/m}$ is the sound velocity in the linear regime.
This means that the low-momentum excitations correspond to phonons.
This linear dispersion relation can also be seen to arise directly by neglecting the second term, which contains the squared derivative of the density. This gives the so-called Luttinger-liquid approximation, which describes the Bose gas as a free phonon gas and captures the most relevant contribution near the ground state.

Finally, let us point out that we consider {\it open boundary conditions} for the phonon fields. The hard walls of a box trap impose a vanishing current through the edges, i.e., conditions of \textit{Neumann type} for the phase, $\partial_z\ph|_{z=0}=\partial_z\ph|_{z=L}=0$. These match typical experimental situations more closely, but lead to a zero mode, namely the global phase. As an alternative, we consider conditions of \textit{Dirichlet type}, in which the phase is pinned at the edges, $\ph|_{z=0}=\ph|_{z=L}=0$. This avoids the zero mode and leads to simpler analytic results.

\section{The discretized model and its dynamics}
\label{app:discretized_model_dynamics}

{\noindent}In this appendix, we collect the technical details of the discretized description used throughout the paper, together with the corresponding normal-mode decomposition and the implementation of the compression dynamics at the covariance-matrix level.

\subsection{Discretized quadratic model}
\label{app:discretized_model}

{\noindent}We consider the coarse-grained lattice fields introduced in \Cref{sec:Discretization},
\begin{equation}
    \Rvec
    =
    \bigl(
    \rholatt_1,\ldots,\rholatt_{\Npxl},
    \phlatt_1,\ldots,\phlatt_{\Npxl}
    \bigr)^T,
\end{equation}
which satisfy the rescaled canonical commutation relations
\begin{equation}
    [\X_j,\X_k]
    =
    i\,\frac{\Omega_{jk}}{\Dz}\,\idop .
\end{equation}
Accordingly, the discretized phononic Hamiltonian takes the quadratic form
\begin{equation}\label{eq:app_H_quadratic_form}
    H
    =
    \frac{\Dz}{2}\,
    \Rvec^{\,T}
    \begin{pmatrix}
        H_{\mathfrak{\rho}} & 0\\[2pt]
        0 & H_{\mathfrak{\phi}}
    \end{pmatrix}
    \Rvec,
\end{equation}
where the two blocks are given by
\begin{equation}\label{eq:app_XP_matrices}
    H_{\mathfrak{\rho}} = g\,\id,
    \qquad
    H_{\mathfrak{\phi}}
    =
    \frac{\mrho}{m\,\Dz^2}\,\mathbb{L}
    +
    \mu\,\id ,
\end{equation}
where $\mu\ge 0$ is an infrared regularization parameter that lifts the zero mode whenever needed.

Notably, the coarse-grained variables are not canonically normalized. If desired, one may introduce canonical lattice quadratures
\begin{equation}
    \widetilde{\rho}_i := \sqrt{\Dz}\,\rholatt_i,
    \qquad
    \widetilde{\phi}_i := \sqrt{\Dz}\,\phlatt_i,
\end{equation}
for which the commutators take the standard form. In the present work, however, it is more convenient to keep the original variables and absorb the factor $1/\Dz$ directly into the symplectic structure and into the Heisenberg equations.

For open boundary conditions of Dirichlet type, one imposes a vanishing phase field outside the two endpoints, $\phlatt_0=\phlatt_{\Npxl+1}=0$,
and the corresponding lattice Laplacian is
\begin{equation}\label{eq:app_Laplacian_D}
  \mathbb L =
   \begin{pmatrix}
    2 & -1 & 0 & \cdots & 0\\
    -1 & 2 & -1 & \ddots & \vdots\\
    0 & -1 & 2 & \ddots & 0\\
    \vdots & \ddots & \ddots & \ddots & -1\\
    0 & \cdots & 0 & -1 & 2
   \end{pmatrix},
\end{equation}
which is diagonalized by the discrete sine transform
\begin{equation}\label{eq:app_DST_Dirichlet}
    O_{jk}
    =
    \sqrt{\frac{2}{\Npxl+1}}\,
    \sin\!\left(
        \frac{\pi j k}{\Npxl+1}
    \right),
    \qquad
    j,k=1,\ldots,\Npxl ,
\end{equation}
with eigenvalues
\begin{equation}\label{eq:app_Laplacian_eigs_D}
    \zeta_k
    =
    4\sin^2\!\left(
        \frac{\pi k}{2(\Npxl+1)}
    \right),
    \qquad
    k=1,\ldots,\Npxl .
\end{equation}
The columns of $O$ are standing waves, with the lowest mode, $k=1$ having no node inside the chain, and each further mode having one more node, up to the highest mode, $k=\Npxl$, which alternates in sign from pixel to pixel. More precisely, the sine profiles obey $(-1)^{j+1}O_{jk}=O_{j,\Npxl+1-k}$, i.e., mode $\Npxl+1-k$ is mode $k$ with alternating signs. In particular, the lowest and highest modes have the same weight on every pixel, which gives $C_{1\Npxl}=1$ in Eq.~\eqref{eq:Dirichlet_extremal_mode_identity}.

For Neumann boundary conditions, the first and last diagonal entries of $\mathbb L$ are $1$ instead of $2$, and $O$ is the discrete cosine transform, $O_{jk}=\sqrt{(2-\delta_{k1})/\Npxl}\,\cos[\pi(k-1)(j-\frac12)/\Npxl]$, with eigenvalues $4\sin^2[\pi(k-1)/(2\Npxl)]$. In that case, for $\mu=0$ the mode $k=1$ is the uniform zero mode.

Since $H_{\mathfrak{\rho}}=g\,\id$, the diagonalization problem reduces entirely to the phase block $H_{\mathfrak{\phi}}$. Considering the orthogonal matrix $O\in \mathrm{O}(\Npxl)$ such that
\begin{equation}\label{eq:app_diag_Hphi}
    O^T H_{\mathfrak{\phi}} O
    =
    {\rm diag}(\nu_1,\ldots,\nu_{\Npxl}) ,
\end{equation}
we obtain the normal-mode quadratures as
\begin{equation}\label{eq:app_normal_modes}
    \boldsymbol{\eta}_{\mathfrak{\rho}} = O^T \boldsymbol{\delta\rho},
    \qquad
    \boldsymbol{\eta}_{\mathfrak{\phi}} = O^T \boldsymbol{\phlatt},
\end{equation}
or, componentwise,
\begin{equation}
    (\normoderho)_k = \sum_{j=1}^{\Npxl} O_{jk}\,\rholatt_j,
    \qquad
    (\normodephi)_k = \sum_{j=1}^{\Npxl} O_{jk}\,\phlatt_j .
\end{equation}
Since $O\oplus O$ is symplectic (also with respect to the rescaled symplectic form), this transformation preserves the canonical structure.
With these definitions, the Hamiltonian decouples into independent harmonic modes,
\begin{equation}\label{eq:app_H_modes}
    H
    =
    \Dz
    \sum_{k=1}^{\Npxl}
    H_k,
    \qquad
    H_k
    =
    \frac12\Bigl(
        g\,(\normoderho)_k^2
        +
        \nu_k\,(\normodephi)_k^2
    \Bigr),
\end{equation}
with single-mode frequencies $\omega_k = \sqrt{g\,\nu_k}.$

Since the Hamiltonian is quadratic and block-diagonal, the covariance matrix of the Gibbs state is also block-diagonal,
\begin{equation}\label{eq:app_Gamma_blockdiag}
    \Gamma_{\tau}
    =
    \begin{pmatrix}
        \Gamma_{\mathfrak{\rho}}(T) & 0\\
        0 & \Gamma_{\mathfrak{\phi}}(T)
    \end{pmatrix}.
\end{equation}
In the normal-mode basis, each mode is an independent harmonic oscillator of frequency $\omega_k$, and the only nonvanishing second moments are given by
\begin{align}\label{eq:app_mode_covariances}
    \aver{(\normoderho)_k^2}_\tau
    &=
    \frac{\omega_k}{2g\,\Dz}\,
    \coth\left(\frac{\omega_k}{2T}\right),
    \\
    \aver{(\normodephi)_k^2}_\tau
    &=
    \frac{g}{2\omega_k\,\Dz}\,
    \coth\left(\frac{\omega_k}{2T}\right).
\end{align}
Equivalently, up to a reordering of the quadratures mode by mode and by the definition in Eq.\,\eqref{eq:CM_def},
\begin{equation}
   S^T \Gamma_{\tau} S
    =
    \bigoplus_{k=1}^{\Npxl}
    \frac{1}{\Dz}
    \begin{pmatrix}
        \dfrac{\omega_k}{g}\coth\bigl(\dfrac{\omega_k}{2T}\bigr) & 0\\[8pt]
        0 & \dfrac{g}{\omega_k}\coth\bigl(\dfrac{\omega_k}{2T}\bigr)
    \end{pmatrix}.
\end{equation}
Transformation back to the lattice basis is achieved again via the symplectic transformation $S=O\oplus O$.

\subsection{Compression dynamics}
\label{app:compression}

{\noindent}We now summarize the effective compression dynamics used in the main text. See also Ref.\,\cite{Gluza_2021} for a discussion on this model.
Let $L(t)$ denote the time-dependent length of the box and define
\begin{equation}
    \lambda(t):=\frac{L(0)}{L(t)},
\end{equation}
so that compression corresponds to an increase in $\lambda(t)$.

In the laboratory frame, the slowly moving boundary induces a hydrodynamic background velocity field. In the long-wavelength approximation, after linearization around the corresponding mean field, the Hamiltonian takes the form~\cite{Gluza_2021}
\begin{equation}\label{eq:app_H_lab_cross}
\begin{aligned}
H(t)=\int_{0}^{L(t)}\!\!\de z\,
\Bigg[&
\frac{\mrho(t)}{2m}\,(\partial_z \ph)^2
+\frac{g}{2}\,\dens^2\\
&+\frac{\dot L(t)\,z}{2L(t)}
\Bigl(
    \dens\,\partial_z\ph
    +
    (\partial_z\ph)\,\dens
\Bigr)
\Bigg].
\end{aligned}
\end{equation}

A convenient description is obtained by passing to the co-compressing coordinate
\begin{equation}
    x=\lambda(t)\,z,
\end{equation}
which keeps the spatial interval fixed, $x\in[0,L]$ with $L=L(0)$.
To preserve the canonical algebra, one must simultaneously rescale the density fluctuation field,
\begin{equation}\label{eq:app_compress_fields_cont}
    \ndd(x,t)
    :=
    \frac{1}{\lambda(t)}
    \dens\!\left(\frac{x}{\lambda(t)},t\right),
    \qquad
    \ph(x,t)
    :=
    \ph\!\left(\frac{x}{\lambda(t)},t\right).
\end{equation}
Then,
\begin{equation}
    [\ndd(x,t),\ph(x',t)] = i\,\delta(x-x'),
\end{equation}
and the Hamiltonian becomes
\begin{equation}\label{eq:app_H_cocompressing}
    H(t)
    =
    \int_0^L \de x\,
    \left[
        \frac{\mrho(0)}{2m}\,\lambda^2(t)\,(\partial_x\ph)^2
        +
        \frac{g}{2}\,\lambda(t)\,\ndd^2
    \right].
\end{equation}
Thus, in the co-compressing frame, the moving-boundary problem is mapped to a quadratic Hamiltonian on a fixed domain with time-dependent couplings.

We then discretize the fixed reference interval into $\Npxl$ pixels and keep $\Npxl$ constant during the protocol.
The instantaneous lattice spacing is therefore
\begin{equation}
    \Dz(t)=\frac{L(t)}{\Npxl}
    =
    \frac{\Dz}{\lambda(t)},
    \qquad
    \Dz\equiv \Dz(0).
\end{equation}
The discrete counterpart of Eq.\,\eqref{eq:app_compress_fields_cont}, obtained by coarse-graining the new field variables, is
\begin{equation}\label{eq:app_nu_i_def_clean}
    \nddisc_i(t)
    :=
    \frac{1}{\lambda(t)}\,\rholatt_i(t),
\end{equation}
while the phases $\phlatt_i(t)$ are not rescaled. The commutators of these variables are time independent:
\begin{equation}
    [\nddisc_j(t),\phlatt_k(t)]
    =
    i\,\frac{\delta_{jk}}{\Dz}.
\end{equation}
In terms of these co-compressing lattice variables, the discretized Hamiltonian reads
\begin{equation}\label{eq:app_H_disc_compress_clean}
H(t)
=
\Dz\left[
\frac{\mrho(0)}{2m}\,\lambda^2(t)
\sum_{i=0}^{\Npxl}
\left(
\frac{\phlatt_{i+1}-\phlatt_i}{\Dz}
\right)^2
+
\frac{g}{2}\,\lambda(t)
\sum_{i=1}^{\Npxl}\nddisc_i^{\,2}
\right],
\end{equation}
with $\phlatt_0=\phlatt_{\Npxl+1}=0$ for Dirichlet boundary conditions, and up to the same zero-mode regularization term as in the static case, if included.
Equivalently,
\begin{equation}\label{eq:app_Hn_blocks}
    H(t)
    =
    \frac{\Dz}{2}\,\mathbf{Y}^{T}(t) \begin{pmatrix}
       H_{\mathfrak{\nddisc}}(t) & 0\\
        0 & H_{\mathfrak{\phlatt}}(t)
    \end{pmatrix} \mathbf{Y}(t) ,
\end{equation}
where $\mathbf{Y}(t) =
    (\nddisc_1(t),\ldots,\nddisc_{\Npxl}(t),\phlatt_1(t),\ldots,\phlatt_{\Npxl}(t))^T$ is the vector of co-compressing pixel variables, and the blocks are given by
\begin{equation}
    H_{\mathfrak{\nddisc}}(t) = g\,\lambda(t)\,\id,
    \qquad
    H_{\mathfrak{\phlatt}}(t)
    =
    \frac{\mrho(0)}{m\,\Dz^2}\,\lambda^2(t)\,\mathbb{L}
    +\mu\,\id .
\end{equation}
Therefore
the model remains a quadratic
model of the same type as in equilibrium, but with time-dependent coefficients.

We implement a finite compression or expansion protocol by splitting the total evolution time $t_{\rm comp}$ into $\Ntro$ short intervals of duration
\begin{equation}
    \delta t = \frac{t_{\rm comp}}{\Ntro}.
\end{equation}
At each step $n=0, \dots, \Ntro-1$, we approximate the dynamics using a piecewise-constant Hamiltonian $H_n$
with length given by the infinitesimal update:
\begin{equation}
    L_{n+1}=(1+\epsilon)L_n,
    \qquad
    \Dz_{n+1}=(1+\epsilon)\Dz_n,
\end{equation}
with $|\epsilon|\ll 1$, so that $\epsilon<0$ corresponds to compression.
Due to the $1/\Dz$ scaling of the commutators, which compensates the prefactor $\Dz$ of the Hamiltonian~\eqref{eq:app_Hn_blocks}, the Heisenberg equation for the phase-space vector $\mathbf{Y}(t)$ takes the form
\begin{equation}\label{eq:app_Heis_generator}
    \frac{d}{dt}\mathbf{Y}(t_n)
    =
    \Omega\big(H^{(n)}_{\mathfrak{\nddisc}}\oplus H^{(n)}_{\mathfrak{\phlatt}}\big)\,\mathbf{Y}(t_n)
    =:
    A_n\,\mathbf{Y}(t_n),
\end{equation}
with the blocks evaluated at the midpoint of the interval, i.e., $H^{(n)}_{\mathfrak{\nddisc}}:=H_{\mathfrak{\nddisc}}(t_n+\delta t/2)$ and $H^{(n)}_{\mathfrak{\phlatt}}:=H_{\mathfrak{\phlatt}}(t_n+\delta t/2)$, which depend on $\lambda_{n+1/2}:=\lambda(t_n+\delta t/2)=(1+\epsilon)^{-(n+1/2)}$. This choice makes the scheme accurate to second order in $\delta t$.
Hence, the exact propagator over one frozen step is the symplectic matrix
\begin{equation}\label{eq:app_Sn_exact_clean}
    S_n
    =
    \exp(A_n\,\delta t) .
\end{equation}
The covariance matrix, expressed in terms of the co-compressing pixel variables $\mathbf{Y}(t_n)$, evolves as
\begin{equation}\label{eq:app_CM_update_clean}
    \tilde \Gamma_{n+1}
    =
    S_n\, \tilde \Gamma_n\,S_n^T ,
\end{equation}
where we used the notation $[\tilde \Gamma_n]_{ij} = \aver{ Y_i  Y_j + Y_j  Y_i} -2\aver{ Y_i}\aver{ Y_j}$ for the instantaneous co-compressing frame covariance matrix.
We choose $\delta t$ such that $\omega_{\max}\delta t\leq0.02$, where $\omega_{\max}$ is the largest mode frequency during the protocol, with at least $10^3$ steps per compression. Halving $\delta t$ changes the witness by less than $10^{-6}$ for all protocols shown in \Cref{fig:compression_regimes}.

At each step $n$, one may further diagonalize the instantaneous phase block $H_{\mathfrak{\phlatt}}^{(n)}$ by the same orthogonal matrix $O$,
\begin{equation}
    O^T H_{\mathfrak{\phlatt}}^{(n)} O
    =
    {\rm diag}(\nu_1^{(n)},\ldots,\nu_{\Npxl}^{(n)}).
\end{equation}
We emphasize that in the present case $O$ is actually independent of the time step $n$: the compression changes only the eigenvalues, not the eigenvectors, because it merely rescales the Laplacian block by the scalar factor $\lambda_{n+1/2}^2$ and the density block by $\lambda_{n+1/2}$.
Thus, for Dirichlet boundary conditions one uses at each time step the DST matrix in Eq.\,\eqref{eq:app_DST_Dirichlet}, while for Neumann boundary conditions, one may use the same discrete cosine transform at every step.

{\bf Remark.---}Note that the passage from the physical density fluctuations $\rholatt_i(t)$ to the co-compressing variables $\nddisc_i(t)$ is local, i.e.,
\begin{equation}
    \mathbf{Y}(t)=R(t)\,\Rvec(t),
    \qquad
    R(t)=\bigl(\lambda^{-1}(t)\id\bigr)\oplus \id .
\end{equation}
At the covariance-matrix level, this gives
\begin{equation}
    \tilde \Gamma(t)=R(t)\,\Gamma(t)\,R^T(t),
    \qquad
    \Gamma(t)=R^{-1}(t)\,\tilde \Gamma(t)\,R^{-T}(t).
\end{equation}
Since $R(t)$ is block diagonal and local in the site partition, this transformation does not create or destroy entanglement across lattice sites.

\section{Gaussian states and entanglement negativity}
\label{app:ent_measures}

{\noindent}Let us consider $\Npxl$ generic bosonic modes, associated with quadratures
\be
{\bf  X} := ( x_1, x_2,\dots, x_{\Npxl}, p_1,
 p_2,\dots, p_{\Npxl})^T
\ee
that obey CCRs given by $[ \X_l, \X_m]=i
\Omega_{l,m}$. For a quantum state $\varrho$ we can take the vector of mean values $\bar{\textbf{X}}:=\aver{\textbf{X}}_{\varrho}=\tr(\varrho\textbf{X})$ and the covariance matrix
\begin{equation}
[\Gamma_\varrho]_{ij}:= \aver{ \X_i  \X_j + \X_j  \X_i}_{\varrho} - 2 \aver{ \X_i}_{\varrho}\aver{ \X_j}_{\varrho}\ ,
\end{equation}
which together fully characterize the state whenever it is Gaussian. See for example \cite{Weedbrook2012,Adesso2014,Serafini2017}.
The single constraint for the covariance matrix to correspond to a physical state is given by the Heisenberg uncertainty relation
\begin{equation}
\Gamma + i\Omega \succeq 0 .
\end{equation}

{\noindent}Besides the covariance-matrix witnesses, which we have discussed earlier, we can discuss other common entanglement quantifiers.
For Gaussian states, the most accessible bipartite measure is the logarithmic negativity, which is obtained from the ``partially transposed'' covariance matrix~\cite{vidalwerner2002,audenaert2002}.
Note that the fact that this measure is a proper entanglement monotone does not follow immediately from its definition, but was proven independently afterwards~\cite{PlenioPRL2005}.
For a bipartition $A|B$ of the $\Npxl$ modes, the partial transpose with respect to $B$ acts in phase space by flipping the sign of the momentum-like quadratures belonging to subsystem $B$.
It is therefore implemented by
\begin{equation}
    \Gamma_{\varrho}^{T_B}
    =
    \Pi_{A|B}\,\Gamma_{\varrho}\,\Pi_{A|B},
    \qquad
    \Pi_{A|B}:=\mathbb{1}_{\Npxl}\oplus \Pi_{p}^{(A|B)},
\end{equation}
where $\Pi_{p}^{(A|B)}$ is diagonal, with entries $+1$ on sites belonging to $A$ and $-1$ on sites belonging to $B$.

Let $\tilde \nu_j$ denote the symplectic eigenvalues of $\Gamma_{\varrho}^{T_B}$, i.e., the positive moduli of the eigenvalues of $i\Omega\Gamma_{\varrho}^{T_B}$; for the rescaled variables of the main text, $[X_j,X_k]=i\Omega_{jk}/\Delta$, the same expressions hold with $\Gamma_\varrho$ replaced by $\Delta\,\Gamma_\varrho$.
Then, the logarithmic negativity across the cut $A|B$ is~\cite{vidalwerner2002,audenaert2002}
\begin{equation}
    E_N(A|B)
    =
    \sum_{j=1}^{\Npxl}
    \max\!\left\{
        0,\,
        -\log_2\!\left(\tilde \nu_j\right)
    \right\}.
    \label{eq:app_logneg_sympl}
\end{equation}

For the Dirichlet thermal state and the alternating cut, $A=\{1,3,\ldots\}$, $B=\{2,4,\ldots\}$, the negativity can be obtained in closed form. The sign pattern $\Pi_p^{(A|B)}={\rm diag}[(-1)^{j+1}]$, i.e., $+1$ on the odd pixels, which belong to $A$, and $-1$ on the even ones, which belong to $B$, obeys $\Pi_p^{(A|B)}O=O\mathcal P$, where $O$ is the discrete sine transform of Eq.\,\eqref{eq:app_DST_Dirichlet} and $\mathcal P$ is the permutation matrix $k\mapsto\Npxl+1-k$ (see \Cref{app:discretized_model}). Hence,
\begin{equation}
    \begin{aligned}
    (O\oplus O)^T\,\Gamma_\tau^{T_B}\,(O\oplus O)&=2\,{\rm diag}\big(\langle(\normoderho)_1^2\rangle_\tau,\ldots,\langle(\normoderho)_{\Npxl}^2\rangle_\tau\big)\\
    &\oplus2\,{\rm diag}\big(\langle(\normodephi)_{\Npxl}^2\rangle_\tau,\ldots,\langle(\normodephi)_1^2\rangle_\tau\big),
    \end{aligned}
    \label{eq:app_PT_Dirichlet}
\end{equation}
i.e., the partial transposition pairs the density quadrature of mode $k$ with the phase quadrature of mode $\Npxl+1-k$, and the symplectic eigenvalues are $\tilde\nu_k=2\Delta\sqrt{\langle(\normoderho)_k^2\rangle_\tau\langle(\normodephi)_{\Npxl+1-k}^2\rangle_\tau}$. Both $\langle(\normoderho)_k^2\rangle_\tau$ and $\langle(\normodephi)_{\Npxl+1-k}^2\rangle_\tau$ increase with $k$, so the smallest symplectic eigenvalue is $\tilde\nu_1=2\Delta\sqrt{\langle(\normoderho)_1^2\rangle_\tau\langle(\normodephi)_{\Npxl}^2\rangle_\tau}=1-\langle W\rangle_\tau$, with $\langle W\rangle_\tau$ given in Eq.~\eqref{eq:product_sep_condition_clean}. Therefore,
\begin{equation}
    E_N(A|B)=\sum_{k=1}^{\Npxl}\max\bigl\{0,-\log_2\tilde\nu_k\bigr\}
\end{equation}
is nonzero if and only if $\langle W\rangle_\tau>0$.

\section{SDP formulation of the covariance matrix criterion}
\label{app:OptSDPwit}

{\noindent}In this appendix, we will state the SDP problems for finding the optimal entanglement witness for determining the inseparability with respect to a given partition into $k$ parties given the covariance matrix of a multi-mode state. In practice, we will be interested in the cases $k=2$, i.e., bipartitions and $k=N$, i.e., full separability. Note that these optimal witnesses are straightforward generalizations of the ones stated explicitly in Ref.\,\cite{hyllus2006optimal} for detecting inseparability in any fixed bipartition.

First of all, let us recall that a \textit{primal} SDP problem takes the form following the same notation of Ref.\,\cite{hyllus2006optimal}:
\begin{align}
    \alpha \coloneqq \min_{\mathbf{x}\,\in\,\mathbb{R}^t} \;\; &\mathbf{c}\cdot\mathbf{x}\\
    \text{subject to}\;\; &F_0+\sum_{i=1}^t F_i x_i \succeq \mathbf{0}\,,\label{eq:primalStd}
\end{align}
where $\mathbf{c}\,\in\,\mathbb{R}^t$ and $\{F_i\}_{i=0}^t$ are Hermitian matrices. The corresponding \textit{dual} problem is given by
\begin{align}
    \beta \coloneqq \max_{Z} \;\; &\!-\tr(F_0 Z)\\
    \text{subject to}\;\; &\,\tr(F_i Z)=c_i\,\text{ for }i=1,\ldots,t\,,\\
    \;\; &\, Z\succeq \mathbf{0}\,.
\end{align}
Note that due to \textit{weak duality} of SDP, the dual optimal solution lower bounds the primal optimal solution (i.e., $\alpha\geq\beta$). Furthermore, \textit{strong duality} holds (i.e., $\alpha=\beta$) whenever \textit{Slater's condition} is fulfilled~\cite{Watrous2018}, i.e., if $\alpha$ is finite (the primal problem is feasible and bounded) and there exists a positive definite matrix $Z\succ \mathbf{0}$ which satisfies all constraints in the dual problem, or, analogously, if $\beta$ is finite and the primal constraint can be satisfied with strict inequality.

To certify inseparability of an $N$-mode state with respect to a fixed $k$-partition of $N$ modes given the state's covariance matrix $\Gamma$, the primal SDP problem corresponding to the optimal witness can be written as:
\begin{align}
    \min_{\gamma^{(1)},\ldots,\gamma^{(k)}, x_e} \;\; &-x_e \label{eq:fixed_k_primal1}\\
    \text{subject to}\;\; &\Gamma-\oplus_{\alpha=1}^k \gamma^{(\alpha)} \succeq \mathbf{0}_{2N}\,,\label{eq:fixed_k-sep1}\\
    &\oplus_{\alpha=1}^k \gamma^{(\alpha)} + (1+x_e)\,i\chi\Omega \succeq \mathbf{0}_{2N}\,,\label{eq:fixed_k-sep2}
\end{align}
since for $\Gamma$ corresponding to a separable state with respect to this $k$-partition, it must satisfy Eqs.\,\eqref{eq:fixed_k-sep1}--\eqref{eq:fixed_k-sep2} (with $\mathbf{0}_{d}$ being a $d\times d$ zero matrix) for $x_e=0$ so that $\gamma^{(\alpha)}$ is a covariance matrix for the $\alpha$-th party of the $N$ modes and $\chi$ is a common rescaling factor of the canonical commutation relation of the conjugate quadratures (e.g., $[\rholatt_{i}, \phlatt_j] = i \delta_{ij} \idop /\Dz$ as described in Sec.\,\ref{sec:Discretization} $\Rightarrow\,\chi=1/\Delta$).
We can combine the two inequality constraints in Eqs.\,\eqref{eq:fixed_k-sep1}--\eqref{eq:fixed_k-sep2} by a direct sum to get one equivalent matrix inequality constraint so that it fits the standard form of the primal SDP in Eq.\,\eqref{eq:primalStd}:
\begin{align}
    &(\Gamma\oplus i\chi\Omega) + \sum_{\alpha=1}^k\sum_{i,j\in\,\mathcal{I}_\alpha,\,i\leq j}(-F_{i,j}\,\oplus \,F_{i,j})\,\tilde{\gamma}_{i,j}^{[k]}\notag\\
    &\qquad+\, x_e(\mathbf{0}_{2N}\oplus i\chi\,\Omega) \,\succeq \mathbf{0}_{4N}\,,\label{eq:fixed_k_primal_combConstraint}
\end{align}
where $\mathcal{I}_\alpha\subset\{1,\ldots,2N\}$ labels the quadratures of the modes that belong to the $\alpha$-th party (i.e., both indices $j$ and $j+N$ for each such mode $j$), $F_{i,j} = \ket{i}\!\!\bra{j}+\ket{j}\!\!\bra{i}$ for $i<j$ and $F_{i,i}=\ket{i}\!\!\bra{i}$, with $i,j\in\{1,\ldots,2N\}$, and $\tilde{\gamma}^{[k]}\coloneqq \oplus_{\alpha=1}^k\gamma^{(\alpha)} = \sum_{i,j=1}^{2N}\tilde{\gamma}_{i,j}^{[k]}\ket{i}\!\!\bra{j}$ is real symmetric, with $\tilde{\gamma}_{i,j}^{[k]}=0$ if $(i,j)\not\in\mathcal{I}_\alpha\times\mathcal{I}_\alpha$ for any $\alpha$. We detect entanglement of $\Gamma$ if $x_e<0$ since this corresponds to the violation of the Heisenberg uncertainty relation which $\oplus_{\alpha=1}^k \gamma^{(\alpha)}$ must fulfill if $\Gamma$ is separable with respect to this $k$-partition.

Due to the direct sum structure of the constraint in Eq.\,\eqref{eq:fixed_k_primal_combConstraint}, the corresponding dual SDP problem is given by
\begin{align}
    \max_{Z_1,Z_2} \;\; &\!-\tr(\Gamma\, Z_1\oplus i\chi\,\Omega\, Z_2)\\
    \text{subject to}\;\; &\tr(i\chi\,\Omega \,Z_2)=-1\,,\\
    &\tr(-F_{i,j}\,Z_1\,\oplus F_{i,j}\,Z_2) = 0 \notag\\
    &\quad\forall\; i,j\in\,\mathcal{I}_\alpha\,,\; \alpha\,\in\,\{1,\ldots,k\}\,,\\
    \;\; &Z_1,Z_2\succeq \mathbf{0}_{2N}\,.
\end{align}
For any Hermitian matrix $Z=\Re(Z)+i\Im(Z)$, it holds that $\Re(Z)^T=\Re(Z)$ and $\Im(Z)^T=-\Im(Z)$. Since $\Gamma$ is real and symmetric, $\tr[\Gamma\,\Im(Z_1)]=0$, and thus
\begin{equation}
    \tr(\Gamma \,Z_1) = \tr\{\Gamma \,[\Re(Z_1)+i\Im(Z_1)]\} = \tr[\Gamma \,\Re(Z_1)]\,,
\end{equation}
and we can simplify the dual problem as follows:
\begin{align}
    \max_{Z_1,Z_2} \;\; &1-\tr(\Gamma \Re(Z_1))\\
    \text{subject to}\;\; &\tr(i\chi\,\Omega \,Z_2)=-1\,,\\
    &\Re(Z_1^{\mathrm{bd}}) = \Re(Z_2^{\mathrm{bd}})\,,\label{eq:fixed_k_dual_eq}\\
    & Z_1,Z_2\succeq \mathbf{0}_{2N}\,,\label{eq:fixed_k_dual_last}
\end{align}
where $Z^{\mathrm{bd}} = \sum_{\alpha=1}^k\sum_{i,j\in\mathcal{I}_\alpha} Z_{i,j}\ket{i}\!\!\bra{j}$ is the result of pinching the matrix $Z$ with respect to the $k$-partition blocks, i.e., of setting to zero all its entries that connect different parties. Following the same argument in Ref.\,\cite{hyllus2006optimal}, the state corresponding to $\Gamma$ is entangled with respect to this fixed $k$-partition if $\tr(\Gamma \,\Re(Z_1))<1$.
\\[-8pt]

Strong duality holds for the SDP problems in Eqs.\,\eqref{eq:fixed_k_primal1}--\eqref{eq:fixed_k_dual_last}, for any multi-mode covariance matrix $\Gamma$. Indeed, we can choose $x_e=-1$ and $\oplus_{\alpha=1}^k \gamma^{(\alpha)} = \mathbf{0}_{2N}$, which is a primal feasible solution. Moreover, Eqs.\,\eqref{eq:fixed_k-sep1}--\eqref{eq:fixed_k-sep2} imply $\Gamma+(1+x_e)\,i\chi\Omega\succeq\mathbf{0}_{2N}$, so that $x_e$ is bounded from above and $\alpha$ is finite. Simultaneously, the positive definite matrices $Z_1=Z_2=a\,\mathbb{1}_{2N}-\frac{i}{2N\chi}\,\Omega$, with $a>\frac{1}{2N\chi}$, satisfy the normalization $\tr(i\chi\,\Omega\,Z_2)=-1$ (since $\tr\,\Omega=0$ and $\Omega^2=-\mathbb{1}_{2N}$, i.e., $\tr\,\Omega^2=-2N$) and the other dual constraint in Eq.\,\eqref{eq:fixed_k_dual_eq}. Therefore, Slater's condition is satisfied and we have strong duality.\\[-8pt]

In this language, detecting non-full-separability is equivalent to detecting ``fixed'' $N$-partition inseparability since there is only one $N$-partition of $N$ modes.
Moreover, we can also observe that the violation of the normalized covariance-based witness, i.e., in particular the result of the dual SDP described above, can be turned into a quantitative lower bound on the so-called best separable approximation (BSA).

\section{CMC witness violation as a lower bound on the best separable approximation}\label{app:bsa}

{\noindent}The \textit{Best Separable Approximation} (BSA) is an entanglement monotone introduced in Refs.\,\cite{LewSanp98,KarnasLew01} and discussed in various subsequent works~\cite{Brandao2005, FadelVitagliano_2021}, also in the context of many-body systems~\cite{mathe2025estimatingentanglementmonotonesnonpure, Quesada_2014, Fr_rot_2023}. It is defined as
\begin{equation}
    E_{\mathrm{BSA}}^{\mathcal S}(\varrho)
    :=
    \inf\Bigl\{
        p\in[0,1]:
        \;
        \varrho=(1-p)\sigma+p\varpi,
    \Bigr\} ,
    \label{eq:app_BSA_def_general}
\end{equation}
where $\sigma\in\mathcal S$ is a separable state and
        $\varpi$ is an arbitrary remainder state.
Thus, $E_{\mathrm{BSA}}^{\mathcal S}(\varrho)$ is the smallest weight of an arbitrary remainder state needed in order to represent $\varrho$ as a convex mixture with a fully separable state.

It is well known that a lower bound on the BSA can be given from properly normalized entanglement witnesses~\cite{Brandao2005, eisert2007quantitative, G_hne_2007, G_hne_2008, Audenaert_2006} as well as covariance-based criteria~\cite{Gittsovich_2010,de_Vicente_2007,Hofmann_2003}. Here, we present a more direct proof for our framework, directly connecting the optimal CMC witness to the BSA.
Assume that we are given a covariance-based witness normalized in the form
\begin{equation}
    W_{\mathcal S}(\Gamma_\varrho)
    :=
    1-\tr\bigl(Z_{\mathcal S}\Gamma_\varrho\bigr),
    \qquad
    Z_{\mathcal S}\succeq 0,
    \label{eq:app_witness_normalized}
\end{equation}
with the property that
\begin{equation}
    \tr\bigl(Z_{\mathcal S}\Gamma_\sigma\bigr)\ge 1
    \qquad
    \forall\,\sigma\in\mathcal S.
    \label{eq:app_witness_sep_bound}
\end{equation}
This is precisely the normalization naturally produced by the dual SDP described above, with $Z_{\mathcal S}=\Re(Z_1)$.
Whenever $W_{\mathcal S}(\Gamma_\varrho)>0$, the state is therefore not contained in $\mathcal S$.

The same witness violation immediately yields a quantitative lower bound on the BSA.
Indeed, let
\begin{equation}
    \varrho=(1-p)\, \sigma\, +p\, \varpi,
    \qquad
    \sigma\in\mathcal S.
\end{equation}
Due to the concavity of the covariance matrix, we have
\begin{equation}
    \Gamma_\varrho \succeq (1-p)\, \Gamma_\sigma\, +\, p\, \Gamma_{\varpi}.
\end{equation}
Since $Z_{\mathcal  S}\succeq0$, it follows that
\begin{equation}
    \tr(Z_{\mathcal  S} \, \Gamma_\varrho)
    \ge
    (1-p)\tr(Z_{\mathcal S}\, \Gamma_\sigma)
    +
    p \,  \tr(Z_{\mathcal  S} \, \Gamma_{\varpi}).
\end{equation}
Using Eq.\,\eqref{eq:app_witness_sep_bound} and $\tr(Z_{\mathcal S}\, \Gamma_{\varpi})\ge0$, we obtain
\begin{equation}
    \tr(Z_{\mathcal S}\, \Gamma_\varrho)\ge 1-p.
\end{equation}
Rearranging gives
\begin{equation}
    p \ge 1-\tr(Z_{\mathcal S}\Gamma_\varrho)
    = W_{\mathcal S}(\Gamma_\varrho).
\end{equation}
Minimizing over all decompositions in Eq.\,\eqref{eq:app_BSA_def_general} yields the witnessed lower bound
\begin{equation}
    E_{\mathrm{BSA}}^{\mathcal S}(\varrho)
    \ge
    \max\!\left\{
        0,\,
        W_{\mathcal S}(\Gamma_\varrho)
    \right\}.
    \label{eq:app_BSA_lowerbound_general}
\end{equation}

\section{Optimal witness for thermal states}
\label{app:thermal_witness_proof}

{\noindent}In this appendix, we prove the results on thermal states summarized in \Cref{sec:LL_optimal_witness_equilibrium}. We first derive the criterion of Eq.~\eqref{eq:general_product_sep_condition}, which is valid for all states. We then show that, for Dirichlet boundary conditions, it is also sufficient for the thermal state, and that the corresponding witness is the optimum of the SDP of \Cref{app:OptSDPwit} at every temperature. The argument has two steps: an inequality that \emph{every} fully separable state must satisfy, and an explicit separable decomposition of the thermal state whenever this inequality is satisfied, which is closely related to the construction of Anders and Winter~\cite{anders2008entanglementseparabilityquantumharmonic}.

\subsection{A criterion valid for all separable states}\label{app:witness_proof}

In the normal-mode basis of \Cref{sec:thermal_state}, the thermal covariance matrix is diagonal,
\begin{equation}
    \begin{aligned}
    S^T\Gamma_\tau S
     =2\operatorname{diag}
       \big(&\langle(\normoderho)_1^2\rangle_\tau,\ldots,\langle(\normoderho)_{\Npxl}^2\rangle_\tau,\\
        &\langle(\normodephi)_1^2\rangle_\tau,\ldots,\langle(\normodephi)_{\Npxl}^2\rangle_\tau\big),
     \quad S:=O\oplus O ,
    \end{aligned}
    \label{eq:thermal_CM_sector_normal_form}
\end{equation}
where the factor two stems from the convention in Eq.~\eqref{eq:CM_def}, and the normal-mode variances are given in Eqs.~\eqref{eq:mode_covariances} and~\eqref{eq:mode_covariances2}. Since $\omega\coth(\omega/2T)$ increases and $\coth(\omega/2T)/\omega$ decreases with $\omega$, and $\omega_k$ grows with $k$, the density variances grow and the phase variances decrease with $k$, as stated in the main text.

To derive Eq.~\eqref{eq:general_product_sep_condition}, choose any real $z_\rho,z_\phi>0$ and take the positive matrix
\begin{equation}
    Z=z_\rho\,\mathbf u\mathbf u^T\oplus z_\phi\,\mathbf w\mathbf w^T ,
    \label{eq:M_two_mode_support}
\end{equation}
i.e., $Z$ projects onto the profile $\mathbf u$ in the density sector and onto the profile $\mathbf w$ in the phase sector. In the normal-mode basis, $Z$ has only two nonzero entries, on the density quadrature of mode $1$ and on the phase quadrature of mode $\Npxl$.
Multiplying such a positive weight matrix with the covariance matrix of a generic state $\varrho$ and taking the trace, we obtain
\begin{equation}
\tr(Z\Gamma_\varrho) = 2z_\rho\operatorname{Var}[(\normoderho)_1]_\varrho +2z_\phi \operatorname{Var}[(\normodephi)_{\Npxl}]_\varrho .
\end{equation}
If $\Gamma_\sigma$ is the
covariance matrix of a fully separable state $\sigma$, the CMC states that there exist physical
single-pixel covariances $\gamma_j$ such that
$\Gamma_\sigma \succeq\bigoplus_j\gamma_j$. We denote by $\gamma_j^{\rho\rho}$ and $\gamma_j^{\phi\phi}$ the diagonal entries of $\gamma_j$, i.e., its (doubled) density and phase variances. Taking the trace with the positive matrix $Z$, we obtain the \emph{linear} bound
\begin{align}
    \tr(Z\Gamma_\sigma)
    &\geq\sum_{j=1}^{\Npxl}
       \left(z_\rho u_j^2\,\gamma_j^{\rho\rho}
              +z_\phi w_j^2\,\gamma_j^{\phi\phi}\right)
       \notag\\
    &\geq\frac{2\sqrt{z_\rho z_\phi}}{\Delta}
       \sum_{j=1}^{\Npxl}|u_jw_j|
     =\frac{2C_{1\Npxl}}{\Delta}\sqrt{z_\rho z_\phi} ,
    \label{eq:linear_extremal_sep_bound}
\end{align}
which must be valid {\it for all separable states} $\sigma$.
In the second line, we used, for every pixel, the inequality between the arithmetic and the geometric mean, $x+y\geq2\sqrt{xy}$ for $x,y\geq0$, with $x=z_\rho u_j^2\gamma_j^{\rho\rho}$ and $y=z_\phi w_j^2\gamma_j^{\phi\phi}$, together with the local uncertainty relation $\gamma_j^{\rho\rho}\gamma_j^{\phi\phi}\geq\Delta^{-2}$, which follows from $\gamma_j+i\Omega/\Delta\succeq0$. Note that the CMC also holds for separable mixtures of
non-Gaussian product states, so Eq.~\eqref{eq:linear_extremal_sep_bound} is
valid without assuming Gaussianity.

It is convenient to normalize this family to have separable bound one. For
$C_{1\Npxl}>0$, take $q>0$ and choose $z_\rho=q\Delta/(2C_{1\Npxl})$ and $z_\phi=\Delta/(2C_{1\Npxl}q)$, so that the right-hand side of Eq.~\eqref{eq:linear_extremal_sep_bound} equals one.
With this choice we get, for every fully separable state, the inequality
\begin{equation}
    \tr(Z_q \Gamma_\sigma)
    =\frac{\Delta}{C_{1\Npxl}}
        \left(q\operatorname{Var}[(\normoderho)_1]_\sigma+\frac{\operatorname{Var}[(\normodephi)_{\Npxl}]_\sigma}{q}\right)\geq1 .
    \label{eq:linear_extremal_family}
\end{equation}
For a given state, the left-hand side of \cref{eq:linear_extremal_family} is smallest at
$q_*=\sqrt{\operatorname{Var}[(\normodephi)_{\Npxl}]/\operatorname{Var}[(\normoderho)_1]}$, i.e., for the weights $z_\rho^*=q_*\Delta/(2C_{1\Npxl})$ and $z_\phi^*=\Delta/(2C_{1\Npxl}q_*)$.
Because the separable inequality holds for \emph{every} $q>0$, we obtain the
equivalent product condition, Eq.~\eqref{eq:general_product_sep_condition}. Violating it certifies entanglement of any state, Gaussian or not, whereas satisfying it does not by itself imply separability.

For Dirichlet boundary conditions, the extremal discrete-sine profiles satisfy
\begin{equation}
    w_j=(-1)^{j+1}u_j,
    \qquad C_{1\Npxl}=\sum_j u_j^2=1,
    \label{eq:Dirichlet_extremal_mode_identity}
\end{equation}
i.e., the highest mode is the lowest one with alternating signs, so that $|u_j|=|w_j|$ on every pixel and $C_{1\Npxl}$ attains its maximal value (see \Cref{app:discretized_model}).
Hence, Eq.~\eqref{eq:general_product_sep_condition}
becomes
\begin{equation}
    \operatorname{Var}[(\normoderho)_1]\,\operatorname{Var}[(\normodephi)_{\Npxl}]
        \geq\frac{1}{4\Delta^2}.
    \label{eq:extremal_mode_conditions}
\end{equation}
For the thermal covariance, substitution of the optimal weights $z_\rho^*$ and $z_\phi^*$ into the linear witness of
Eq.~\eqref{eq:linear_witness_rewrite} yields
\begin{equation}
    \langle W\rangle_\tau
     =1-\tr(Z_{\rm D}\Gamma_\tau)
     =1-2\Delta\sqrt{
        \langle(\normoderho)_1^2\rangle_\tau
        \langle(\normodephi)_{\Npxl}^2\rangle_\tau},
    \label{eq:product_sep_condition_clean}
\end{equation}
where $Z_{\rm D}$ denotes the matrix of Eq.~\eqref{eq:M_two_mode_support} with these weights and $C_{1\Npxl}=1$, and we used that the thermal first moments vanish. Inserting the thermal variances of Eqs.~\eqref{eq:mode_covariances} and~\eqref{eq:mode_covariances2} gives the second form of Eq.~\eqref{eq:optwit_thermal_analytic}.

\subsection{An explicit separable decomposition of the thermal state}

We now show that, for Dirichlet boundary conditions, this witness gives the exact separability threshold of the thermal state. Following Ref.~\cite{anders2008entanglementseparabilityquantumharmonic}, consider the product state with covariance $\Gamma_0(q)/\Delta$, i.e., the same pure, uncorrelated state on every pixel, where
\begin{equation}
    \Gamma_0(q):=
       q^{-1}\mathbb{1}_{\Npxl}\oplus q\,\mathbb{1}_{\Npxl},
    \qquad q>0,
    \label{eq:Gamma0_LL_def_clean}
\end{equation}
written for the ordering $(\rholatt_1,\ldots,\rholatt_{\Npxl},\phlatt_1,\ldots,\phlatt_{\Npxl})$ of the quadratures. Reordering them pixel by pixel, as $(\rholatt_1,\phlatt_1,\ldots,\rholatt_{\Npxl},\phlatt_{\Npxl})$, with the permutation matrix $P$ whose only nonzero entries are $P_{2j-1,j}=P_{2j,\Npxl+j}=1$, for $j=1,\ldots,\Npxl$, gives
\begin{equation}
    P\,\Gamma_0(q)\,P^T=\bigoplus_{j=1}^{\Npxl}
         \begin{pmatrix}q^{-1}&0\\0&q\end{pmatrix},
    \label{eq:Gamma0_pixelwise}
\end{equation}
which shows explicitly that $\Gamma_0(q)$ is the same uncorrelated covariance on every pixel. The permutation $P$ is only a relabeling of the quadratures, which maps the symplectic form $\Omega$ onto $\bigoplus_j\left(\begin{smallmatrix}0&1\\-1&0\end{smallmatrix}\right)$.

The matrix $\Gamma_0(q)/\Delta$ is a physical
product covariance and thus provides an explicit separability certificate for the thermal state
if and only if it fits below the full covariance,
\begin{equation}
    \Gamma_\tau-\frac{1}{\Delta}\Gamma_0(q)\succeq0.
    \label{eq:Gamma_minus_Gamma0_psd}
\end{equation}
This condition is easy to check because $\Gamma_0(q)$ is invariant under
$S=O \oplus O$, while $S^T\Gamma_\tau S$ is diagonal, see Eq.~\eqref{eq:thermal_CM_sector_normal_form}. Since the density variances $\langle(\normoderho)_k^2\rangle_\tau$ increase and the phase variances $\langle(\normodephi)_k^2\rangle_\tau$ decrease with $k$, Eq.~\eqref{eq:Gamma_minus_Gamma0_psd}
reduces to just the two extremal conditions $2\langle(\normoderho)_1^2\rangle_\tau\geq1/(q\Delta)$ and $2\langle(\normodephi)_{\Npxl}^2\rangle_\tau\geq q/\Delta$. Both conditions can be met by some $q$ precisely when $\langle(\normoderho)_1^2\rangle_\tau\langle(\normodephi)_{\Npxl}^2\rangle_\tau\geq1/(4\Delta^2)$ and a suitable choice is precisely $q_*=\sqrt{\langle(\normodephi)_{\Npxl}^2\rangle_\tau/\langle(\normoderho)_1^2\rangle_\tau}$, the same value that optimizes the witness in Eq.~\eqref{eq:linear_extremal_family}.
The difference in Eq.~\eqref{eq:Gamma_minus_Gamma0_psd} is then explicitly
positive semidefinite in \emph{all} normal modes.
Together with
Eq.~\eqref{eq:extremal_mode_conditions}, this proves that for Dirichlet boundary conditions
\begin{equation}
    \begin{aligned}
       \langle(\normoderho)_1^2\rangle_\tau\,\langle(\normodephi)_{\Npxl}^2\rangle_\tau<\frac{1}{4\Delta^2}
           &\quad\Longrightarrow\quad\tau\text{ is entangled},\\
       \langle(\normoderho)_1^2\rangle_\tau\,\langle(\normodephi)_{\Npxl}^2\rangle_\tau\geq\frac{1}{4\Delta^2}
           &\quad\Longrightarrow\quad\tau\text{ is fully separable}.
    \end{aligned}
    \label{eq:Dirichlet_thermal_complete_criterion}
\end{equation}
The first implication holds for arbitrary states (with the second moments replaced by variances, see Eq.~\eqref{eq:extremal_mode_conditions}), whereas the second one uses the concrete thermal covariance and Gaussianity.

\subsection{Other boundary conditions}

For other boundary conditions, the overlap $C_{1\Npxl}$ need not equal one, e.g., $C_{1\Npxl}\to2\sqrt2/\pi\approx0.90$ for the Neumann chain at large $\Npxl$.
The two simple constructions then give distinct rigorous bounds,
\begin{equation}
    \begin{aligned}
       &\operatorname{Var}[(\normoderho)_1]\,\operatorname{Var}[(\normodephi)_{\Npxl}]
           <\frac{C_{1\Npxl}^{\,2}}{4\Delta^{2}}
           \ \Longrightarrow\ \text{entanglement},\\
       &\langle(\normoderho)_1^2\rangle_\tau\,\langle(\normodephi)_{\Npxl}^2\rangle_\tau\geq\frac{1}{4\Delta^{2}}
           \ \Longrightarrow\
             \begin{array}{l}\text{Gaussian thermal}\\ \text{full separability}.\end{array}
    \end{aligned}
    \label{eq:general_boundary_two_sided_bounds}
\end{equation}
The interval between these thresholds requires the unrestricted CMC SDP. In particular, the Dirichlet two-mode formula does not correspond to the exact SDP optimum for (regularized) Neumann boundary conditions.

\subsection{Relation to bipartite entanglement}

For any state and any bipartition $A|B$, one has $\langle W_{\rm opt}\rangle\geq\langle W_{\rm opt}^{A|B}\rangle\geq1-\tilde\nu_{\min}^{A|B}$, where $W_{\rm opt}^{A|B}$ is the optimal CMC witness for that bipartition: the first inequality holds because full separability implies separability across $A|B$, the second because partial transposition maps any decomposition $\Gamma\succeq\gamma_A\oplus\gamma_B$ with (relaxed) physical local covariances onto an analogous decomposition of $\Gamma^{T_B}$, which then obeys the same (relaxed) uncertainty relation. For the Dirichlet thermal state and the alternating cut, the smallest symplectic eigenvalue of the partially transposed covariance is $\tilde\nu_{\min}^{A|B}=1-\langle W(T)\rangle$ (\Cref{app:ent_measures}), so that all three quantities coincide.

\section{Scaling of the optimal witness with physical parameters}
\label{app:analytic_scaling}

{\noindent}For Dirichlet boundary conditions, the normal-mode frequencies are $\omega_k=2(c_0/\Delta)\sin[\pi k/(2(\Npxl+1))]$, where $c_0=\sqrt{g\mrho/m}$ is the speed of sound. With $\theta_{\Npxl}:=\pi/[2(\Npxl+1)]$, so that $\omega_1=2(c_0/\Delta)\sin\theta_{\Npxl}$ and $\omega_{\Npxl}=2(c_0/\Delta)\cos\theta_{\Npxl}$, the witness of Eq.~\eqref{eq:optwit_thermal_analytic} becomes
\begin{equation}
    \begin{aligned}
    W(T)
    =
    1&-
    \sqrt{\tan\theta_{\Npxl}}\\
    &\times\sqrt{
    \coth\left(\frac{c_0}{T \Delta}\sin\theta_{\Npxl}\right)
    \coth\left(\frac{c_0}{T \Delta}\cos\theta_{\Npxl}\right)
    }.
    \end{aligned}
    \label{eq:app_W_theta}
\end{equation}
At zero temperature, both hyperbolic cotangents equal one, and $W(0)=1-\sqrt{\tan\theta_{\Npxl}}$ depends only on the number of pixels.

\subsection{Separability threshold in temperature}
\label{app:threshold}

{\noindent}The lowest frequency, $\omega_1=2(c_0/\Delta)\sin\theta_{\Npxl}\simeq\pi c_0/L$, is typically much smaller than the relevant temperatures. For $T\gg\omega_1$, which at $T\simeq T_{\rm sep}$ requires $\Npxl\gg1$, the first hyperbolic cotangent in Eq.~\eqref{eq:app_W_theta} can be replaced by $T\Delta/(c_0\sin\theta_{\Npxl})$, and the factors $\sin\theta_{\Npxl}$ cancel, giving
\begin{equation}
    W(T)\simeq1-\sqrt{\frac{\coth y}{y}} ,
    \qquad
    y:=\frac{c_0}{T\Delta}\cos\theta_{\Npxl}=\frac{\omega_{\Npxl}}{2T} .
    \label{eq:app_W_highT}
\end{equation}
The witness thus depends only on the ratio between the temperature and the highest mode frequency. It is positive if and only if $y>y_*$, where $y_*\simeq1.1997$ is the positive solution of $\coth y_*=y_*$. The separability temperature is therefore
\begin{equation}
    T_{\rm sep}\simeq\frac{\omega_{\Npxl}}{2y_*}\simeq0.417\,\omega_{\Npxl}\simeq0.834\,\frac{c_0}{\Delta}\cos\theta_{\Npxl} ,
    \label{eq:app_Tsep}
\end{equation}
or, restoring physical units and using $\hbar c_0/\heal=g\mrho$, $k_BT_{\rm sep}\simeq0.83\,\hbar c_0/\Delta=0.83\,(\heal/\Delta)\,g\mrho$ for $\Npxl\gg1$. Remarkably, $T_{\rm sep}$ does not depend on the system size $L$, but only on the coarse-graining scale: finer partitions remain entangled up to higher temperatures, $T_{\rm sep}\propto1/\Delta$. For $T\ll\omega_{\Npxl}$, i.e., $y\gg1$, Eq.~\eqref{eq:app_W_highT} further simplifies to $W(T)\simeq1-y^{-1/2}\simeq1-\sqrt{T\Delta/c_0}$, which is the same scaling as for a continuum limit and fixed length, as we observe below.

\subsection{Continuum limit at fixed physical length} \label{app:continuum_limit}

{\noindent}In the continuum limit, $\Npxl\to\infty$ and $\Delta=L/\Npxl\to0$ at fixed $L$, the argument of the first hyperbolic cotangent in Eq.~\eqref{eq:app_W_theta} tends to $x_L:=\pi c_0/(2LT)$, that of the second one diverges, and $\tan\theta_{\Npxl}\simeq\pi\Delta/(2L)$, so that
\begin{equation}
    W(T)\simeq1-\sqrt{\frac{\pi\Delta}{2L}\,\coth\left(\frac{\pi c_0}{2LT}\right)}\;\longrightarrow\;1 ,
    \label{eq:app_W_fixedL_Dz}
\end{equation}
with a leading correction of order $\sqrt{\Delta}$, which becomes $W(T)\simeq1-\sqrt{T\Delta/c_0}$ for large $L$, i.e., for $x_L\ll1$.

In the formal continuum limit at fixed $T$ and $L$, the witness thus saturates, $W\to1$.
Physically, however, $\Delta$ cannot be taken much below the healing length, where the phonon description breaks down (see \Cref{app:bogoliubovtheory}), and both the value of the witness and the threshold temperature of Eq.~\eqref{eq:app_Tsep} should be interpreted within this range. Larger pixels, $\Delta\gtrsim\heal$, are instead fully within the range of the phonon description and are relevant for experiments with a finite imaging resolution, e.g., $T_{\rm sep}\simeq18$\,nK for $\Delta=4\,\heal\simeq1\,\mu$m (\Cref{fig:ent_over_temp}).

\section{Optimal CMC witness transported during compression}

\subsection{Invariance under local squeezing}
\label{app:local_squeezing_invariance}

The optimal value of the SDP does not change under local symplectic transformations $S_{\rm loc}$, which act separately on each pixel, such as the transformation $D_\lambda$ of Eq.~\eqref{eq:adiabatic_covariance_local_squeezing} that describes adiabatic compression. Indeed, if $\Gamma\succeq\bigoplus_j\gamma_j$ with $\gamma_j+is\Omega/\Delta\succeq0$ for some $s>0$ (i.e., $s=1+x_e$ in the primal SDP of \Cref{app:OptSDPwit}), then $S_{\rm loc}\Gamma S_{\rm loc}^T\succeq\bigoplus_jS_j\gamma_jS_j^T$, where $S_j$ is the action of $S_{\rm loc}$ on pixel $j$, and $S_j\gamma_jS_j^T+is\Omega/\Delta\succeq0$, since $S_j\Omega S_j^T=\Omega$. The converse follows in the same way with $S_{\rm loc}^{-1}$. Correspondingly, a witness matrix $Z$ that obeys $\tr(Z\Gamma_\sigma)\geq1$ for all fully separable states is mapped onto $S_{\rm loc}^{-T}ZS_{\rm loc}^{-1}$, which obeys the same bound and has the same value on the transformed state. Hence, the adiabatically compressed state has the same optimal witness value as the initial thermal state, and the thermal two-mode witness, transported by $D_\lambda$, is optimal for it. We also verified this numerically, by solving the SDP for the adiabatically compressed state.
\subsection{Bound on the entanglement created by compression}
\label{app:compression_bound}

{\noindent}Here we derive Eq.~\eqref{eq:compression_bound} for
a compression in which $\lambda(t)$ grows monotonically. As in our numerical scheme (\Cref{app:compression}), we describe the compression as a sequence of small sudden steps, each followed by a free evolution with the Hamiltonian held fixed, and we can see that the following bound holds
\begin{equation}
    \tilde\Gamma(t)\succeq\lambda^{-1/2}(t)\,\tilde\Gamma_{\rm ad}(t) ,
    \label{eq:app_covariance_bound}
\end{equation}
where $\tilde\Gamma_{\rm ad}(t)$ is the covariance matrix of the adiabatic state in Eq.~\eqref{eq:adiabatic_covariance_local_squeezing}. At $t=0$, both sides are equal to $\Gamma_\tau$. Let us now consider a {\it sudden} infinitesimal step from $\lambda$ to $\lambda+\de\lambda$, with $\de\lambda>0$, and assume that Eq.~\eqref{eq:app_covariance_bound} holds before it.
The step does not change $\tilde\Gamma$, since the state has no time to evolve (while the Hamiltonian has changed according to the compression factor). On the other hand, in an infinitesimal {\it adiabatic} step with the same compression factor, the covariance matrix follows the Hamiltonian eigenmodes and is given by Eq.~\eqref{eq:adiabatic_covariance_local_squeezing}, i.e., $\tilde\Gamma_{\rm ad} =(\lambda+\de\lambda)^{1/2}\cdot \Gamma_{\mathfrak{\rho}}(T)\oplus (\lambda+\de\lambda)^{-1/2}\cdot \Gamma_{\mathfrak{\phi}}(T)$, so that the right-hand side of Eq.~\eqref{eq:app_covariance_bound} becomes $\Gamma_{\mathfrak{\rho}}(T)\oplus(\lambda+\de\lambda)^{-1}\,\Gamma_{\mathfrak{\phi}}(T)\preceq\Gamma_{\mathfrak{\rho}}(T)\oplus\lambda^{-1}\,\Gamma_{\mathfrak{\phi}}(T)$, i.e., it cannot exceed its value before the step. Hence we have that the bound in \cref{eq:app_covariance_bound} still holds. During the following free evolution, $\tilde\Gamma\to S_n\tilde\Gamma S_n^T$, as in Eq.~\eqref{eq:app_CM_update_clean}, which preserves the inequality and leaves $\tilde\Gamma_{\rm ad}$ unchanged, because $\tilde\Gamma_{\rm ad}$ is the covariance matrix of a Gibbs state of the Hamiltonian $H_n$ that generates $S_n$ (\Cref{sec:LL_optimal_witness_adiabatic}). Repeating this reasoning for an infinite sequence of infinitesimal steps thus leads to the same bound holding also in the limit of a continuous compression.

For any witness matrix $Z\succeq0$ normalized such that $\tr(Z\Gamma_\sigma)\geq1$ for all fully separable states $\sigma$, Eq.~\eqref{eq:app_covariance_bound} gives
\begin{equation}
    \tr[Z\tilde\Gamma(t)]\geq\lambda^{-1/2}(t)\,\tr[Z\tilde\Gamma_{\rm ad}(t)]\geq\lambda^{-1/2}(t)\,[1-\langle W(T)\rangle] ,
\end{equation}
where we used that the optimal witness of the adiabatic state equals that of the initial thermal state (\Cref{app:local_squeezing_invariance}). Maximizing $1-\tr[Z\tilde\Gamma(t)]$ over $Z$ yields Eq.~\eqref{eq:compression_bound}. As long as the free evolution preserves Eq.~\eqref{eq:app_covariance_bound}, as in our case, the bound also holds after the compression.

Combined with the high-temperature form of the witness, $\langle W\rangle\simeq1-\sqrt{\coth(y)/y}$ (\Cref{app:threshold}), the bound shows that a compression up to $\lambda_f$ can only create entanglement if the initial temperature is below $T_{\rm act}$, defined by $\coth(y)/y=\lambda_f$ with $y=\omega_{\Npxl}/(2T_{\rm act})$. Using $\coth y\simeq1/y+y/3$, this gives $T_{\rm act}\simeq1.2\,(\lambda_f-1/3)^{1/2}\,T_{\rm sep}$.
The bound is approached shortly after a sudden compression. Such a compression does not change the covariance matrix, so that every density variance is $\lambda_f^{-1/2}$ times its adiabatic value. A quarter of an oscillation period of the highest mode later, its phase variance is also $\lambda_f^{-1/2}$ times its adiabatic value, while the lowest mode has barely moved, since $\omega_1\ll\omega_{\Npxl}$. For $\Npxl\gg1$, the two-mode witness then approaches the bound.

\section{Open-system dynamics and steady state with local and global master equations}
\label{app:thermal_after_comp}

{\noindent}During the contact with the bath, the length of the gas, and thus the compression factor $\lambda_f$, is fixed, so that the Hamiltonian $H_f=\frac{\Delta}{2}\mathbf Y^T\mathbf H_f\mathbf Y$, with $\mathbf H_f=g\lambda_f\id\oplus\frac{\mrho(0)}{m\Delta^2}\lambda_f^2\mathbb L$ (\Cref{app:compression}), does not depend on time. Its normal modes are those of \Cref{sec:thermal_state}, with frequencies $\omega_k^f=\lambda_f^{3/2}\omega_k$. The dynamics of the gas coupled to a thermal bath at temperature $T_b$ is governed by a GKSL master equation, $\de\varrho/\de t=-i[H_f,\varrho]+\mathcal{L}_{\mathrm{diss}}\varrho$. Because $H_f$ is quadratic and the jump operators in $\mathcal{L}_{\mathrm{diss}}$ are linear in the field operators, the dynamics is Gaussian, and the master equation reduces to the Lyapunov equation $\dot\Gamma=G\Gamma+\Gamma G^T+D$ of \Cref{sec:thermalization_after_compression}~\cite{Toscano_2022, De_Chiara_2018}.

\subsection{Global master equation}

{\noindent}When the internal lattice interactions (the Laplacian in $H_f$) are strong compared to the system-bath coupling, the environment resolves the energy levels of the whole gas. The global GKSL master equation is therefore expressed in terms of the normal-mode annihilation and creation operators, $\hat b_k$ and $\hat b_k^\dagger$, which diagonalize $H_f$~\cite{Toscano_2022, Hofer_2017}:
\begin{equation}
    \begin{aligned}
    \frac{\de\varrho}{\de t} = -i[H_f, \varrho] &+ \sum_{k=1}^{N_p} \gamma \Big[ (\bar{n}_k + 1)\mathcal{D}[\hat b_k]\varrho\\
    &+ \bar{n}_k\mathcal{D}[\hat b_k^\dagger]\varrho \Big],
    \end{aligned}
\end{equation}
where $\mathcal{D}[\hat c]\varrho = \hat c\varrho \hat c^\dagger - \frac{1}{2}\{\hat c^\dagger \hat c,\varrho\}$ for a jump operator $\hat c$, $\gamma$ is the uniform system-bath coupling rate, and $\bar{n}_k = [\exp(\omega_k^f/T_b) - 1]^{-1}$ is the Bose-Einstein equilibrium occupation. Since the dissipator shares the eigenbasis of the Hamiltonian and its rates obey detailed balance at the temperature $T_b$, the steady state is the Gibbs state of $H_f$ at temperature $T_b$. In phase space, the drift matrix is $G=\Omega\mathbf H_f-\frac{\gamma}{2}\id$, and the diffusion matrix is $D_{\rm glob}=\gamma\Gamma_\infty^{\rm glob}$~\cite{Toscano_2022}, where $\Gamma_\infty^{\rm glob}$ is the Gibbs covariance of Eq.~\eqref{eq:app_mode_covariances}, with $g\to g\lambda_f$, $\nu_k\to\lambda_f^2\nu_k$, and $T\to T_b$. This steady state is in fact unique: in the normal-mode basis, $G$ block-diagonalizes with eigenvalues
$-\gamma/2\pm i\omega_k^f$, so that, for any $\gamma>0$, every deviation from the steady state decays, and $\Gamma(t)\to\Gamma_\infty^{\rm glob}$ from any initial condition at a rate $\gamma$ \cite{Weedbrook2012}. The global master equation is justified when the system-bath
coupling is weak compared to the level spacing of $H_f$, $\gamma \ll \min_{k\neq k'}|\omega_k^f-\omega_{k'}^f|$. In this regime, the standard Born-Markov-secular
derivation applies directly in the eigenbasis of $H_f$, since coherences between distinct normal
modes oscillate too fast for the bath to resolve, leaving only the population-damping terms retained above \cite{Hofer_2017}.

\subsection{Local master equation}

{\noindent}In phenomenological models, dissipation is often modeled locally, assuming independent thermal baths couple directly to each spatially localized site $j$, ignoring the internal interaction Laplacian. We define the local ladder operators $\hat{a}_j = \sqrt{\frac{\Delta}{2}}(\ell^{1/2}\hat{\xi}_j + i\ell^{-1/2}\hat{\phi}_j)$, which obey $[\hat a_j,\hat a_j^\dagger]=1$. The length $\ell$ is needed because $\xi_j$ has dimensions of an inverse length, and it fixes the local reference oscillator toward which the bath drives each pixel. The local GKSL master equation then reads
\begin{equation}
    \begin{aligned}
    \frac{\de\varrho}{\de t} = -i[H_f, \varrho] &+ \sum_{j=1}^{N_p} \gamma \Big[ (\bar{n}_{\rm loc} + 1)\mathcal{D}[\hat{a}_j]\varrho+ \bar{n}_{\rm loc} \mathcal{D}[\hat{a}_j^\dagger]\varrho \Big],
    \end{aligned}
\end{equation}
with $\bar n_{\rm loc}=[\exp(\omega_{\rm loc}/T_b)-1]^{-1}$. As reference oscillator, we choose the on-site part of $H_f$, $\frac{\Delta}{2}(g\lambda_f\,\xi_j^2+s_0\,\phi_j^2)$, with $s_0=2\mrho(0)\lambda_f^2/(m\Delta^2)$ for Dirichlet boundary conditions. It corresponds to $\ell=\sqrt{g\lambda_f/s_0}$, and its frequency, $\omega_{\rm loc}=\sqrt{g\lambda_f s_0}\simeq\omega_{\Npxl}^f/\sqrt2$ for many pixels, sets $\bar n_{\rm loc}$. The drift matrix $G$ is the same as for the global master equation, whereas the diffusion matrix $D_{\rm loc}$ of \Cref{sec:thermalization_after_compression} is diagonal and the same on every pixel. It is thus not changed by the orthogonal transformation $O\oplus O$ to the normal modes, so that the steady-state equation $G\Gamma_\infty^{\rm loc}+\Gamma_\infty^{\rm loc}G^T+D_{\rm loc}=0$ can be solved separately for each mode (local and global master equations for coupled harmonic oscillators are compared, e.g., in Refs.~\cite{Hofer_2017,De_Chiara_2018,Gonzalez_2017}). With $\tilde r:=g\lambda_f/\ell$ and $\tilde s_k:=\lambda_f^2\nu_k\ell$, so that $\tilde r\tilde s_k=(\omega_k^f)^2$, the steady state of mode $k$ is given by
\begin{equation}
    \begin{aligned}
    2\langle(\normoderho)_k^2\rangle_\infty&=\frac{\vartheta_{\rm loc}}{\Delta\ell}\,\frac{\gamma^2+2\tilde s_k(\tilde r+\tilde s_k)}{\gamma^2+4(\omega_k^f)^2},\\
    2\langle(\normodephi)_k^2\rangle_\infty&=\frac{\vartheta_{\rm loc}\,\ell}{\Delta}\,\frac{\gamma^2+2\tilde r(\tilde r+\tilde s_k)}{\gamma^2+4(\omega_k^f)^2},\\
    \langle\{(\normoderho)_k,(\normodephi)_k\}\rangle_\infty&=\frac{\vartheta_{\rm loc}}{\Delta}\,\frac{\gamma\,(\tilde s_k-\tilde r)}{\gamma^2+4(\omega_k^f)^2},
    \end{aligned}
    \label{eq:app_local_steady_state}
\end{equation}
where $\vartheta_{\rm loc}=2\bar n_{\rm loc}+1$, while all correlations between different modes vanish. In general, the steady state of each mode thus contains correlations between its density and its phase, so that it is not a thermal state. These correlations vanish, however, for a normal mode that matches the reference oscillator, $\tilde s_k=\tilde r$, and for all modes for weak damping, $\gamma\ll\omega_1^f$, which is the regime that we consider. In this regime, each mode relaxes to a diagonal, thermal-like state, whose density variances grow and phase variances decrease with $k$, as in a thermal state, so that the analysis of \Cref{app:thermal_witness_proof} can be applied again.

Regarding the validity of this dynamics, a local master equation can be derived microscopically if the coupling between neighboring sites is weak compared with the damping rate, which in turn has to be small compared with the frequencies of the sites, as required by the weak-coupling (Born--Markov) approximation~\cite{Hofer_2017,Gonzalez_2017,De_Chiara_2018}. In our chain, however, the coupling between neighboring pixels, given by the off-diagonal entries of the Laplacian in $H_f$, is half of the on-site coefficient $s_0$, i.e., comparable to the frequency $\omega_{\rm loc}$ of a single pixel, so that there is no regime of $\gamma$ in which both conditions hold. We therefore treat the local master equation as a phenomenological model, and we only consider weak damping, $\gamma\ll\omega_1^f$, as required by the weak-coupling approximation. Even in this regime, its steady state is not the Gibbs state of $H_f$, as shown above~\cite{De_Chiara_2018}.

\bibliography{bibliography}

\end{document}